\documentclass[12pt]{article}
\usepackage[utf8]{inputenc}

\usepackage[authoryear]{natbib}

\usepackage{amsmath}
\usepackage{xspace}
\usepackage{amssymb}
\usepackage{graphicx}
\usepackage{lineno}
\usepackage{setspace}
\usepackage{natbib}
\usepackage{rotating}
\usepackage{mwe}
\usepackage{pdflscape}

\newcommand{\appref}[1]{\mbox{Appendix~\ref{#1}}}
\newcommand{\secref}[1]{\mbox{\S$\,$\ref{#1}}}
\newcommand{\figref}[1]{\mbox{Figure~\ref{#1}}}
\newcommand{\tabref}[1]{\mbox{Table~\ref{#1}}}

\newcommand{\cd}{\,|\,}

\newcommand{\evid}{{\cal E}}
\newcommand{\kns}{{\cal K}}
\newcommand{\ukns}{{\cal U}}
\newcommand{\yloci}{{\cal Y}}
\newcommand{\hap}{{\cal H}}
\newcommand{\dbase}{{\cal D}}
\newcommand{\regd}{\textsuperscript{\textregistered}}

\newcommand{\tvec}[1]{{\underline{#1}}}

\title{A sub-critical branching process model for application to analysing Y~haplotype DNA mixtures}

\author{Robert G.  Cowell\\(Retired, without affiliation)}

\begin{document}
\maketitle

\begin{abstract}

The treatment of short-tandem-repeat (STR) loci  on the Y chromosome presents special problems in the forensic analysis of DNA mixtures, chiefly 
but not exclusively relating to the linkage of Y-STR loci which 
precludes the use of the `product rule' for estimating Y-haplotype match probabilities. 
In  recent paper, \cite{andersen2017convincing} estimated, via a population simulation model, the  distribution of the number of haplotypes sharing a common profile over a set of Y-STR loci, and argued for its use as an alternative to estimating Y-haplotype match probabilities. 

In this paper we present a sub-critical branching process model that approximates their population model, and show how to estimate the haplotype number distribution numerically using multivariate probability generating functions. It is shown that the approximation provides a good fit to their simulations.
The model is extended to propose a new framework  for evaluating the weight-of-evidence of Y-STR haplotype mixtures, and it is illustrated with publicly available data of a three person DNA mixture.


\end{abstract}

\section{Introduction}
\label{sec:introduction}

Although the processes involved in the PCR amplification of short-tandem-repeat  loci in the Y chromosome (Y-STR loci) are the same as for autosomal loci, it has long been acknowledged that 
the linkage of the Y-STR loci on the Y-chromosome means that the product rule for evaluating profile probabilities of people  used for 
autosomal loci cannot be applied to Y haplotype profiles. This has created difficulties for developing and implementing statistical  models for the evaluation of likelihood ratios of  DNA profile evidence on Y-STR loci. Apart from acknowledging the problems posed by Y-STR profiles, there is little consensus amongst forensic scientists on the way forward in the evaluation of statistical evidence for presentation in court.

\cite{andersen2017convincing} question whether it is reasonable to attach a match probability to a Y-STR haplotype profile. They propose as an alternative estimating the 
number of matching males in a population using a simulation that takes into account the population growth rate and the mutation rates  of the Y-STR loci.

In this paper the population growth model of \citep{andersen2017convincing} is approximated by a sub-critical branching process model for the distinct haplotype profiles. It is shown how to find the haplotype profile multiplicity distribution using multivariate probability generating functions. This is applied, with an extension, to a novel framework for estimating the weight-of-evidence of Y-STR haplotype profiles of DNA
samples, and is illustrated with an application to a three person mixture. 

The plan of the paper is as follows. In \secref{sec:challenge} provides more background  of the challenges in estimating Y-STR haplotype probabilities that led to this paper. 
In \secref{sec:subcriticalbp} the \cite{andersen2017convincing} model is recast as a sub-critical branching process model using multivariate probability generating functions.
In \secref{sec:framework} we propose new framework for 
evaluating the weight-of-evidence of Y haplotype profiles in DNA mixtures, with a specific match probability model based on the material in  \secref{sec:subcriticalbp}
described in \secref{sec:match} 
The analysis of a publicly available three person Y-STR mixture data is presented in \secref{sec:realdata}.

\section{Challenges in estimating a Y-STR haplotype profile probability}
\label{sec:challenge}

For readers unfamiliar with the application area, a comprehensive description of the preparation and analysis of forensic DNA samples may be found in \citep{butler2011advanced} and \citep{butler2014advanced}.

\subsection{Profile probability estimation}
\label{sec:ystrprob}

The major challenge of attaching a match probability to a given Y-STR haplotype arises from the lack of recombination, leading to the expectation of a strong correlation between loci, thus rendering the product rule inappropriate. 
Many papers consider the problem of estimating the haplotype match probability  of a person of interest whose haplotype has not been seen in a reference database.
Simple estimates based on 
database counts  provide rather poor estimates \citep{buckleton2011interpretation}, because most haplotypes appear only once in a population database, and the haplotype of a person of interest is unlikely to appear in a database. Brenner's kappa method \citep{brenner2010fundamental,brenner2014understanding} is an adjustment of the database counting method taking into account the rarity of haplotypes, but is limited to singletons. A coalescent based model was proposed by \cite{andersen2013estimating}, but is computationally impractical for forensic casework.
\cite{caliebe2015no} looked at finding quasi-independent subsets of loci  using entropy estimates, but concluded that the dependence was too complicated to be used for finding match probabilities.
Perhaps the most widely used model is the discrete Laplace model \citep{andersen2013discrete}, however it is restricted  to alleles that are
whole numbers of repeats, and cannot deal with alleles having partial repeats or with multicopy loci. An alternative method based on Chow-Liu trees \citep{andersen2018modelling} can lift this limitation. It gives similar results to the discrete Laplace model, however both appear to give probabilities that are too small. The tree based model tries to cater for loci dependence using cross-entropy between loci.  \cite{taylor2018likelihood}  introduce the \textit{haplotype centred} method for evaluating likelihood ratios.
It uses a mutation model, allowing likelihoods to be evaluated locus-by-locus. The model reduces to  the kappa method for single source, unambiguous haplotypes. However it requires having the haplotype of  a person of interest, and cannot cope with partial repeats or locus duplications. It is not clear how 
the  approach would work for deconvolution of a mixture if there are no persons of interest available.

\cite{andersen2017convincing}  proposed  a simulation model  to estimate the number of matching males in a population given the number of loci and mutation rates on each. Their simulations show that the estimates are robust to variation in the population growth rate and variance in reproductive rates. They suggest that these count estimates, rather than match probabilities, should be presented in court. Their model does not take account of partial repeats or duplications. Nor does it take account of any specific details of the observed alleles in the Y-STR loci of the haplotype. This is in line with the suggestion of \cite{brenner2014understanding} that detailed allelic information of  haplotype structure
is not of high importance for profile probability estimation.

\subsection{Profiles with partial repeats}
\label{sec:repeats}

Although most Y haplotype STR locus profiles have alleles described by whole integer repeats, a small minority have partial repeats. Such partial repeats are not a problem for autosomal loci, however they are a problem for Y haplotype profile probability models that employ mutation models such as \cite{andersen2017convincing,andersen2013discrete,taylor2018likelihood}. The discrete Laplace model in particular is based upon the Laplace probability distribution over integers. Partial repeats can be readily incorporated into the tree-based model of \cite{andersen2018modelling}, however their rarity may lead to poor estimation. 

\subsection{Deletions and duplications of loci}
\label{sec:deldups}

Regions on the Y-chromosome can be deleted or duplicated \citep{butler2005chromosomal,butler2011advanced,butler2014advanced}.
 Thus, if a contributor to a DNA sample has a Y-chromosome with a region in which one or more of the Y-STR loci are deleted, then they will not contribute to any allelic peaks on those loci seen in the electropherogram. Alternatively, if loci are duplicated, then the allelic peaks will be boosted by the extra copies of the loci.
Additionally, two or more instances of a duplicated locus can have different alleles so that a person can have two or more alleles on a locus  profile.
 Clearly such artefacts have to be catered for by models using peak height information. The linkage of the loci on the Y chromosome implies that if one
locus is deleted or duplicated, then that increases the possibility that other nearby loci are also deleted or duplicated; that is, locus deletions and duplications tend to be correlated between neighbouring loci. This is an important  factor that also needs to be addressed; it is hard to see how models that attempt to find haplotype profile probabilities using a locus-by-locus probability formula will cope with such correlations.

\section{The  haplotype multiplicity branching process model.}
\label{sec:subcriticalbp}
In the simulation model of \cite{andersen2017convincing},
an initial population of haplotypes is generated on a set of loci (of integer-valued repeat alleles) of known mutation rates, and this
population is evolved with mutation over many generations with some given population growth rate per generation. 
From the final three generations (representing the live males) a sample of haplotypes is taken, and for each  the number of matching live male haplotypes is found. From these values, the distribution of matching haplotypes is estimated, which can be used to estimate the mean number or an upper quantile of matching haplotypes. The authors find that this number does not depend on the population size and that it is relatively insensitive to the assumptions of the reproductive distribution of each generation, provided the size of the population is large.  The main determinants are the number of loci, their mutation rates, and population growth rate.

Their results can be understood as an example of a \textit{sub-critical branching process} as follows. If the population growth rate is denoted by $\gamma$, then the expected number of offspring of a haplotype in each generation is $1+\gamma$. If the loci have mutation rates per generation denoted by $\mu_i$ for the $i^{th}$ locus, then the probability that a given haplotype offspring is identical to a given parent is the probability of no mutations in any locus, given by
$\prod_i (1-\mu_i)$. The expected number of haplotype offspring identical to the parent is therefore
$(1+\gamma)\prod_i (1-\mu_i)$. With  many loci, this will typically be less than one and,  as demonstrated by \cite{brenner2014understanding}, it is extremely remote for a descendant   
haplotype which has  several locus mutations away from a given founder, to have offspring in future generations that  will mutate to a haplotype having the same profile as the original founder. Hence, to a very good approximation,  we have a sub-critical branching process in which any haplotype that appears will eventually die out. 

In the particular case that the number of offspring per generation has a Poisson distribution (which for a large population is equivalent to the Wright-Fisher model), then the total number of haplotypes identical to the parent haplotype will have a Poisson distribution with rate
$\lambda = (1+\gamma)\prod_i (1-\mu_i)$. 
For a sub-critical Poisson branching process the total number $n$ of the founder haplotype plus the number of its identical  descendants will have a Borel distribution \citep{borel1942emploi} with $P(n) = e^{-\lambda n} (\lambda n)^{n-1}/n!$, which has finite mean $1/(1-\lambda)$.
However, it is not the distribution of the total number of descendants, but instead the distribution of the number of identical haplotypes in the final generation and one or two generations back (this approximately covering the age-ranges of living males in an evolving population, as argued in \citep{andersen2017convincing}). A model for evaluating such a distribution, under the assumption of a Poisson distributed number of offspring,  is developed in the next few subsections.

\subsection{Population model.}

Our Y-haplotype population model is similar to a Wright-Fisher model, in which generations do not overlap, but we allow the population to grow in size. Specifically, we model the offspring of a given haplotype as a Galton-Watson branching process in which the number of offspring has a Poisson distribution with mean $\lambda$, hence  the overall growth rate per generation of the population has mean $\lambda$. 

We are interested in  Y-haplotypes profiles over some  large set of 
Y-STR loci $\cal Y$, (typically 20-30 in forensic applications). Let the overall mutation rate per generation of these loci be denoted by $\mu$; we assume that $\lambda(1-\mu) < 1$. We shall assume an infinite-sites model approximation for the profiles on these loci. This means that if a haplotype produces a mutated offspring, the haplotype profile of that offspring has never
occurred before in the population, and is also different to all other haplotypes of its generation. The approximation is justified by the arguments of \citep{brenner2014understanding} as described above.

Under these assumptions, for a particular haplotype, the expected number of offspring that have the same profile as the parent is $\lambda(1-\mu)$, which is less than 1, and so the distribution of descendants of  a particular haplotype profile retaining the same profile is a sub-critical branching process: eventually the \textit{particular} haplotype profile disappears from the population, to be replaced by other haplotype profiles provided the whole population does not die out. 

Suppose that the population has been growing for a large number of generations. We group the haplotypes in the current generation into equivalence classes, in which two haplotypes are in the same class if and only if they have the same profile on all the loci $\cal Y$. We shall use the term cluster to refer to an  equivalence class of haplotypes.  Let $c_j$ denote the number of clusters that have size $j$. Associate the (probability generating function marker) variable $t_j$ to clusters of size $j$, and let
$\tvec{t} = (t_1,t_2,\ldots)$ denote the collection of such variables, such that 
 the multivariate
probability generating function (PGF) for the joint distribution of cluster sizes  $\tvec{C} = (C_1, C_2, \ldots)$ may be written as
\begin{equation}
G(\tvec{t}) = \sum_{\tvec{c}} p(\tvec{c}) t_1^{c_1} t_2^{c_2} t_3^{c_3} \cdots . \label{eq:genfun1}
\end{equation}
If $N$ denotes the size of the population (in the current generation), then 
\begin{equation}
N = \sum_j jc_j, \label{eq:popsize}
\end{equation}
and the summation in \eqref{eq:genfun1} is over all sets of integers
$\tvec{c} = \{c_1,c_2,\ldots\}$ obeying the  constraint \eqref{eq:popsize}. 
For a fixed set of integers $\tvec{c}$, let the number of clusters be denoted by $C$, that is,
$$ C = \sum_j c_j , $$
and denote the proportion of clusters of size $j$ by 
$$ f_j = \frac{c_j}{C}.$$

From the usual properties of PGFs, we have that
$G(\tvec{t}= \tvec{1}) = \sum_{\tvec{c}} p(\tvec{c}) = 1$. We also have that the expected number of clusters of size $k$, denoted by $E[C_k]$, may be found by forming the partial derivative of $G(\underline{t})$ with respect to $t_k$, and then setting all of the $t_j = 1$:
$$ E[C_k] = \frac{\partial G(\tvec{t})}{\partial t_k}
\bigg\vert_{\tvec{t} = \tvec{1}}
= \sum_{\tvec{c}} p(\tvec{c})c_k = \sum_{c_k} p_k(c_k)c_k,$$
where $p_k(c_k)$ is the marginal probability for the number of clusters of size $k$ to be equal to $c_k$.

Higher order moments may be found in a similar manner using higher order derivatives, but we shall not require these.

We now consider the distribution of cluster sizes in the next generation. We  break down the branching process generation of the new offspring into two steps: first generate the offspring (ignoring mutation) and then apply possible mutation to the generated offspring.

In one generation, an individual generates a Poisson($\lambda$) distributed number of offspring. Consider a cluster of $j$ identical haplotypes, represented by $t_j$. The number of offspring generated by all of the haplotypes in the cluster will have a Poisson($j\lambda$)
distribution. If we ignore mutation, the PGF for the
cluster sizes of offspring of such a cluster is:
\begin{equation}
t_j \to e^{-j\lambda}\sum_{k=0}^\infty \frac{ (j\lambda)^k}{k!} t_k .
\label{eq:nomut1}
\end{equation}
If we now include mutation, each of the individual haplotypes has a probability of $\mu$  mutating to a novel haplotype, or not mutating and sharing the same profile as the parent with probability $1-\mu$.
For an offspring  cluster of $k$ such haplotypes, represented by $t_k$ in \eqref{eq:nomut1}, the PGF for such clusters of mutated and non-mutated haplotypes is therefore
\begin{equation}
t_k \to  \sum_{m=0}^k t_1^m t_{k-m} \binom{k}{m} \mu^m(1-\mu)^{k-m}.
\label{eq:mut1}
\end{equation}
Note that both the $m=k$ term and  $m = k-1$ term are multiples of $t_1^k$, which will be important when forming partial derivatives.

Thus to obtain the PGF for the cluster sizes in the next generation,
we substitute for $t_k$ in \eqref{eq:nomut1}
the expression on the right-hand-side of \eqref{eq:mut1}, and in turn
substitute the resulting expression of $t_j$ into the PGF 
of \eqref{eq:genfun1}. To this end, denote the composition of \eqref{eq:nomut1}
and \eqref{eq:mut1} by
\begin{equation}
\tau_j(\tvec{t}) = 
e^{-j\lambda}\sum_{k=0}^\infty \frac{ (j\lambda)^k}{k!}
 \sum_{m=0}^k t_1^m t_{k-m} \binom{k}{m} \mu^m(1-\mu)^{k-m}.
 \label{eq:tauj}
\end{equation}

The PGF for the distribution of cluster sizes in the next generation is then given by $G(\underline{\tau}(\underline{t}))$, which  may be differentiated with respect to the $\underline{t}$ using \eqref{eq:genfun1} and \eqref{eq:tauj} by the chain rule.
Note that $\tau_j(\underline{1}) = 1$.

\subsection{Moments of next generation.}
\label{sec:moment1gen}

We find mean of the number of clusters of each size in the next generation by 
differentiation of the PGF. Denoting the number of clusters of size $r$ in the next generation
by $C^*_r$ we have
$$ E[C_r^*] = \frac{\partial  G(\underline{\tau}(\underline{t}))}{\partial  t_r}\bigg\vert_{ \underline{t}=\underline{1}}.$$

By the chain rule, this will be
\begin{align}
 E[C_r^*] &= \sum_j\frac{\partial  G(\tvec{\tau}(\tvec{t}))}{\partial  \tau_j} \frac{\partial  \tau_j}{\partial  t_r}\bigg\vert_{ \underline{t}=\tvec{1}} \nonumber \\
 &= \sum_j (\sum_{c_j} p_j(c_j)c_j) \frac{\partial  \tau_j}{\partial  t_r}\bigg\vert_{ \tvec{t}=\tvec{1}} \nonumber \\ 
  &= \sum_j E[C_j] \frac{\partial  \tau_j}{\partial  t_r}\bigg\vert_{ \tvec{t}=\tvec{1}} .
 \label{eq:chrule1}
\end{align}

To evaluate the derivative in \eqref{eq:chrule1} we need to treat $t_1$ differently to  $t_j$ with $j>1$, because of the remark after
\eqref{eq:mut1}.

\subsubsection{Differentiation with respect to $t_1$.}

From \eqref{eq:tauj},
\begin{align*}
\frac{\partial \tau_j}{\partial t_1}\bigg\vert_{\tvec{t}=\tvec{1}}
&= 
e^{-j\lambda}\sum_{k=0}^\infty \frac{ (j\lambda)^k}{k!} 
\left(
 \sum_{m=0}^k mt_1^{m-1} t_{k-m} \binom{k}{m} \mu^m(1-\mu)^{k-m} + kt_1^{k-1}\mu^{k-1}(1-\mu)
\right)\bigg\vert_{t=1}\\
&=
e^{-j\lambda}\sum_{k=0}^\infty \frac{ (j\lambda)^k}{k!} 
\left(
k\mu + k(1-\mu)\mu^{k-1}
\right)  . \\
\end{align*}
We obtain the second terms in the inner brackets because the $m=k-1$   term of 
$$ \sum_{m=0}^k t_1^m t_{k-m} \binom{k}{m} \mu^m(1-\mu)^{k-m}$$
is equal to $k\mu^{k-1}(1-\mu) t_1^k$.
It follows that
\begin{align}
\frac{\partial \tau_j}{\partial t_1}\bigg\vert_{\tvec{t}=\tvec{1}}
&= 
e^{-j\lambda}\sum_{k=0}^\infty \frac{ (j\lambda)^k}{k!} 
\left(
k\mu + k(1-\mu)\mu^{k-1}
\right) \nonumber \\
&= 
 e^{-j\lambda}j\lambda\mu \times \left( \sum_{k=1}^\infty \frac{ (j\lambda)^{k-1}}{(k-1)!} \right)
 +
 e^{-j\lambda} j\lambda(1-\mu)\times \left( \sum_{k=1}^\infty \frac{ (j\lambda\mu)^{k-1}}{(k-1)!} \right) \nonumber
 \\
 &= j\lambda\mu  + j\lambda(1-\mu)e^{-j\lambda(1-\mu)}.
 \label{eq:dtuaj1}
 \end{align}

\subsubsection{Differentiation with respect to $t_m$ with $m>1$.}

For convenience we rewrite \eqref{eq:tauj} as

$$\tau_j = e^{-j\lambda}\sum_{k=0}^\infty \frac{ (j\lambda)^k}{k!}  \sum_{m=0}^k t_1^{k-m} t_{m} \binom{k}{m} (1-\mu)^m\mu^{k-m}.$$

We now differentiate with respect to $t_m$ where $m > 1$:

\begin{align}
\frac{\partial \tau_j}{ \partial t_m}\bigg\vert_{\tvec{t}=\tvec{1}}
&= e^{-j\lambda}\sum_{k=0}^\infty  \frac{ (j\lambda)^k}{k!} 
\binom{k}{m} (1-\mu)^m\mu^{k-m} \nonumber
\\
&=
e^{-j\lambda}
\frac{(j\lambda(1-\mu))^m}{m!}\sum_{k=m}^\infty \frac{ (j\lambda\mu)^{k-m}}{(k-m)!}  \nonumber\\
&= e^{-j\lambda}
\frac{(j\lambda(1-\mu))^m}{m!} e^{j\lambda\mu}  \nonumber\\
&= 
\frac{(j\lambda(1-\mu))^m}{m!}e^{-j\lambda(1-\mu)}
\label{eq:dtuajm}
\end{align}

Note that the right-hand-side of \eqref{eq:dtuajm} is the probability $P(X=m)$ where $X\sim\mbox{Poisson}(j\lambda(1-\mu))$.

\subsubsection{Putting it together.}

Using \eqref{eq:dtuaj1} and \eqref{eq:dtuajm} with  \eqref{eq:chrule1} we obtain
\begin{align}
E[C_1^*] &= \sum_j \left(j\lambda\mu  + j\lambda(1-\mu)e^{-j\lambda(1-\mu)}\right) E[C_j] \label{eq:iter1}\\
E[C_m^*] &= \sum_j \frac{(j\lambda(1-\mu))^m}{m!}e^{-j\lambda(1-\mu)} E[C_j] &\mbox{for } m > 1\label{eq:iter2}
\end{align}

We now make the \textit{assumption} that 
the proportion of clusters of each given size $j$ remains, to a very high  approximation, constant
from generation to generation. An argument for a justification of this
assumption may be found in \appref{sec:converge}.
For the current generation, denote these the  cluster size equilibrium probabilities by
$$ \tilde{f}_k = \frac{E[C_k]}{\sum_k E[C_k]},$$ 
and for the next generation by
$$ \tilde{f}_k^* = \frac{E[C_k^*]}{\sum_k E[C_k^*]} \,\, .$$
Under the equilibrium assumption, 
$ \tilde{f}_k = \tilde{f}_k^* $.
Let $C = \sum_k E[C_k]$ and $C^* = \sum_k E[C^*_k]$. Then \eqref{eq:iter1} and \eqref{eq:iter2} become:

\begin{align}
 \tilde{f}_1^* C^* &= 	C \sum_j \left(j\lambda\mu  + j\lambda(1-\mu)e^{-j\lambda(1-\mu)}\right) \tilde{f}_j ,\label{eq:iter3}\\
\tilde{f}_m^*C^* &=  	C \sum_j \frac{(j\lambda(1-\mu))^m}{m!}e^{-j\lambda(1-\mu)} \tilde{f}_j &\mbox{for } m > 1 . \label{eq:iter4}
\end{align}

To numerically solve this coupled set of equations for the equilibrium
probabilities $\tilde{f}_k = \tilde{f}_k^*$  we may proceed as follows.
Initialize $\tilde{f}_1 = 1$, 
$\tilde{f}_m  = 0$ for $m > 1$, and set $C = 1$.
Then iteratively: 
\begin{itemize}
\item substitute the  values of the $\tilde{f}_k$ into the right-hand-sides of  \eqref{eq:iter3} and \eqref{eq:iter4} to obtain the set of  $\tilde{f}_k^*C^*$ values; 
\item normalize these values to sum to 1 to obtain a new set of  values for  the  $\tilde{f}_k$, that is, set
$$\tilde{f}_k := \frac{\tilde{f}_k^*C^*}{\sum_k \tilde{f}_k^*C^*} = \frac{\tilde{f}_k^*}{\sum_k \tilde{f}_k^*},$$
\end{itemize}
until  convergence in the distribution of sum-normalized values is reached to the desired tolerance. It is of course necessary to fix an upper bound on the number of equations $m$ to carry out this iterative scheme. 

\subsection{Including more than the current generation}

\cite{andersen2017convincing} 
defined the population of live males in their simulations as the final three
generations of haplotypes in a  simulated population.
It is straightforward, but a little tedious, to include these extra two generations in the analysis above. For convenience we refer to these final successive generations as the current generation, the next generation and the second generation (the latter is thus the final generation).  We begin by combining the haplotypes from two successive generations

\subsection{Cluster size distribution for two successive generations combined.}

Our starting point is the PGF of the joint distribution of cluster sizes of the current generation, $G(\tvec{t})$ of \eqref{eq:genfun1}.
We consider the offspring of this generation, and construct the PGF for the distribution of cluster sizes in both generations, which we will denote by $G_2(\tvec{t})$.

As for the analysis of the previous sections,  we break the formation of the offspring generation down into two steps: (i) producing the offspring, and (ii) applying random mutation to the offspring. The number of  offspring of haplotypes in a cluster of size $j$ will have a Poisson($j\lambda$) distribution. as before. The total number of such haplotypes in both generations combined will be described by
\begin{equation}
t_j \to e^{-j\lambda}\sum_{k=0}^\infty \frac{ (j\lambda)^k}{k!} t_{j+k} .\label{eq:gen2c}
\end{equation}
What is different from the previous single generation analysis, \eqref{eq:nomut1},  is that the last term is $t_{j+k}$ rather than
$t_k$. The $k$ represents the $k$ children in the next generation, so adding $j$ to the subscript means that the haplotypes from the original cluster are also counted, as they all have the same haplotype if there is no mutation. 

Under mutation we will have that the term
\begin{equation}
t_{k+j} \to  \sum_{m=0}^k t_1^m t_{j+k-m} \binom{k}{m} \mu^m(1-\mu)^{k-m} .\label{eq:gen2d}
\end{equation}
which takes account that only the $k$ children can possibly mutate to a new (unique)  haplotype, the haplotypes of the $j$ parents are fixed. If $m$ of the $k$ do mutate,  they generate $m$ unique haplotypes, represented by the term $t_1^m$. The remaining $k-m$ that do not mutate have haplotypes identical to those in the original cluster of $j$ haplotypes, making a new cluster of size $j+k-m$, represented by the $t_{j+k-m}$ term.
Defining
\begin{equation}
\tau_j(\tvec{t}) =
e^{-j\lambda}\sum_{k=0}^\infty \frac{ (j\lambda)^k}{k!} 
\sum_{m=0}^k t_1^m t_{j+k-m} \binom{k}{m} \mu^m(1-\mu)^{k-m} ,
\label{eq:gen2e}
\end{equation}
the PGF of the joint distribution of cluster sizes in both generations  is given by
\begin{align}
G_2(\tvec{t}) &= \sum_{\tvec{c}^*} p(\tvec{c}^*)t^{c_1^*}
t^{c_2^*} t^{c_3^*} \cdots \nonumber \\
&=G(\tvec{\tau}(\tvec{t})) \nonumber \\
&= \sum_{\tvec{c}} p(\tvec{c})\tau_1(\tvec{t})^{c_1}
\tau_2(\tvec{t})^{c_2}
\tau_3(\tvec{t})^{c_3} \cdots ,\label{eq:gen2f}
\end{align}
where now the $\tvec{C}^*$ are the cluster size numbers in the combined generations, and $C^* = \sum_k C^*_k = \sum_k E[C^*_k]$.

We now differentiate \eqref{eq:gen2f} to obtain expected values of cluster sizes in the combined generations.
\begin{align}
E[C_r^*] &= \frac{\partial  G_2(\tvec{t})}{\partial  t_r} \bigg\vert_{ \underline{t}=\tvec{1}} \nonumber \\
&= \sum_j\frac{\partial  G(\tvec{\tau}(\tvec{t}))}{\partial  \tau_j} \frac{\partial  \tau_j}{\partial  t_r}\bigg\vert_{ \underline{t}=\tvec{1}} \nonumber \\
 &= \sum_j (\sum_{c_j} p_j(c_j)c_j)\frac{\partial  \tau_j}{\partial  t_r}\bigg\vert_{ \tvec{t}=\tvec{1}} \nonumber \\
 &= \sum_j E[C_j]\frac{\partial  \tau_j}{\partial  t_r}\bigg\vert_{ \tvec{t}=\tvec{1}}  
 \label{eq:chrule3}
\end{align}

If the current generation has reached equilibrium with regard to its distribution of cluster sizes, the combined generations will also  have reached equilibrium with regard to its distribution of cluster sizes, so that
\begin{equation}
E[C_r^*] = C^* \tilde{f}^*_r  
=C \sum_j \tilde{f}_j\frac{\partial  \tau_j}{\partial  t_r}\bigg\vert_{ \tvec{t}=\tvec{1}}.  \label{eq:chrule3b}
\end{equation}

Thus to find the probabilities $\tilde{f}^*_r$, we substitute the equilibrium
cluster proportion probabilities $\tilde{f}_j$ obtained for the current generation into \eqref{eq:chrule3b}, using $C=1$, and sum-normalize the set of
values $C^* \tilde{f}^*_r$.  Note that no iterations of this are  required (and in general we would not expect $\tilde{f}^*_r = \tilde{f}_r$).

As before, the evaluation of the 
partial derivatives requires evaluating the
partial derivative of  $t_1$ as a special case.

\subsubsection{Differentiating with respect to $t_1$}

In the terms in \eqref{eq:gen2e}, the term $t_{j+k-m}$ can equal $t_1$ only if $j=0$ or $j=1$. In these cases, if  $j=1$ then  $m=k$, or if $j=0$ then  $m=k-1$. However we only consider 
$j>0$ (i.e., a haplotype has a parent) hence only the $j=1$ case will have extra terms.
If we  denote partial differentiation with respect to $t_z$ by $D_z$ then for $j=1$ we have
\begin{align}
D_1 \tau_1(\tvec{t}) \vert_{t=1}
&=e^{-\lambda}\sum_{k=0}^\infty \frac{ (\lambda)^k}{k!} 
 \left(
 \sum_{m=0}^k m t_1^{m-1} t_{1+k-m} \binom{k}{m} \mu^m(1-\mu)^{k-m}
 + t_1^k \mu^k \right)_{t=1}  \nonumber \\
&= e^{-\lambda}\sum_{k=0}^\infty \frac{ (\lambda)^k}{k!} 
(k\mu + \mu^k) \nonumber \\
&= \lambda\mu + e^{-\lambda(1-\mu)}, \label{eq:gen2g}
\end{align}
and for $j>1$ we have
\begin{align}
D_1 \tau_j \vert_{t=1}
&=e^{-j\lambda}\sum_{k=0}^\infty \frac{ (j\lambda)^k}{k!} 
 \left(
 \sum_{m=0}^k m t_1^{m-1} t_{j+k-m} \binom{k}{m} \mu^m(1-\mu)^{k-m} 
\right)_{t=1} \nonumber\\
&= e^{-j\lambda}\sum_{k=0}^\infty \frac{ (j\lambda)^k}{k!} (k\mu  )
 \nonumber \\
&= j\lambda\mu.\label{eq:gen2h}
\end{align}

\subsubsection{Differentiating with respect to $t_z$ with $z>1$.}
We rewrite \eqref{eq:gen2e} in the equivalent form
\begin{equation}
\tau_j(\tvec{t}) = e^{-j\lambda}\sum_{k=0}^\infty \frac{ (j\lambda)^k}{k!}  \sum_{m=0}^k t_1^{k-m} t_{j+m} \binom{k}{m} (1-\mu)^m\mu^{k-m}.
\label{eq:gen2i}
\end{equation}
Hence for $j+m >1 $:
\begin{align*}
D_{j+m} \tau_j(\tvec{t})\vert_{t=1} 
&=  e^{-j\lambda}\sum_{k=0}^\infty \frac{ (j\lambda)^k}{k!} 
 t_1^{k-m}  \binom{k}{m} (1-\mu)^m\mu^{k-m} \vert_{t=1} \\
&=
e^{-j\lambda} \frac{ (j\lambda(1-\mu))^m}{m!}
\sum_{k=0}^\infty \frac{ (j\lambda\mu)^{k-m}}{(k-m)!}\\
&= 
\frac{(j\lambda(1-\mu))^m}{m!}e^{-j\lambda(1-\mu)},
\end{align*}
that is, with $z = j+m$ so that $z > 1$
\begin{equation}
D_{z} \tau_j(\tvec{t})\vert_{t=1} 
=
\frac{(j\lambda(1-\mu))^{z-j}}{(z-j)!}e^{-j\lambda(1-\mu)}.
\label{eq:gen2j}
\end{equation}

\subsubsection{Putting is all together}
Combining \eqref{eq:chrule3}, \eqref{eq:gen2g}, \eqref{eq:gen2h} and 
 \eqref{eq:gen2j} we obtain
\begin{align*}
C^* \tilde{f}_1^* = E[C_1^*] &=  C \left[(\lambda\mu  + e^{-\lambda(1-\mu)})\tilde{f}_1 + 
 \sum_{j=2} j\lambda\mu \tilde{f}_j\right],\\
C^* \tilde{f}_m^* =E[C_m^*] &= C \sum_{j=1}^m \frac{(j\lambda(1-\mu))^{m-j}}{(m-j)!}e^{-j\lambda(1-\mu)} \tilde{f}_j .
\end{align*}

Note that the coefficient of $\tilde{f}_j$ is the probability 
$P(X = m-j)$ where $X \sim\mbox{Poisson}(j\lambda(1-\mu))$.
Setting $C=1$ and substituting the  $\tilde{f}_1$, normalising the $C^* \tilde{f}_k^*$ gives the  equilibrium cluster size proportions $\tilde{f}_k^*$ for the two combined generations.

\subsection{Two generations back}

The procedure for combining the cluster distributions of three successive generations is similar to the previous subsection which had two successive generations, however the calculation is more involved.

The starting point is again an equilibrium  distribution of cluster sizes of a single generation, with $t_j$ denoting clusters of size $j$ in its PGF.

Then for the next generation we have the substitutions of the previous subsection, for both generations combined.
\begin{equation}
t_j \to e^{-j\lambda}\sum_{k=0}^\infty \frac{ (j\lambda)^k}{k!}  \sum_{m=0}^k t_1^{k-m} t_{j+m} \binom{k}{m} (1-\mu)^m\mu^{k-m}.
\label{eq:firstgen}
\end{equation}

This represents the $j$ identical haplotypes having $k$ offspring, of which 
$k-m$ mutate to produce $k-m$ different singleton haplotypes, exactly as  before in \eqref{eq:gen2e}.

We now allow the first generation offspring to produce offspring for the second  generation. Each of the $k-m$ singletons represented by $t_1^{k-m}$ in \eqref{eq:firstgen}
are in this generation and therfore
can produce offspring, and $m$ of the haplotypes in the $t_{j+m}$ cluster can. In addition these second generation haplotypes can themselves be mutated versions of their first generation parents.
First  
change $t \to s$ in \eqref{eq:firstgen}, that is  write
\begin{equation}
\tau_j(\tvec{s}) = e^{-j\lambda}\sum_{k=0}^\infty \frac{ (j\lambda)^k}{k!}  \sum_{m=0}^k s_1^{k-m} s_{j+m} \binom{k}{m} (1-\mu)^m\mu^{k-m}.
\label{eq:gen3a}
\end{equation}
The $s$'s have themselves to be substituted to represent the production of the second generation. 

Let $b(x;n,\mu) = \binom{n}{x}\mu^x(1-\mu)^{n-x}$ denote the Binomial$(n,\mu)$ distribution probabilities.
On rewriting \eqref{eq:gen3a} using the $b(x;n,\mu)$
we  obtain 

\begin{equation}
\tau_j(\tvec{s}) = e^{-j\lambda}\sum_{k=0}^\infty \frac{ (j\lambda)^k}{k!}  \sum_{m=0}^k b(k-m;k,\mu) s_1^{k-m} s_{j+m} .
\label{eq:gen3b}
\end{equation}

All the $s_1$ can multiply for the next generation, but only $m$ of the $s_{j+m}$ can. Explicitly, adding another (second) generation we have

\begin{align}
\tau_j(\tvec{s}) &=  e^{-j\lambda}\sum_{k=0}^\infty \frac{ (j\lambda)^k}{k!}  \sum_{m=0}^k b(k-m;k,\mu) s_1^{k-m} s_{j+m} \nonumber \\
&\to \nonumber\\
\tau_j(\tvec{t}) &= e^{-j\lambda}\sum_{k=0}^\infty \frac{ (j\lambda)^k}{k!}  \sum_{m=0}^k b(k-m;k,\mu) \nonumber \\&
\times \left(
e^{-\lambda}\sum_{i=0}^\infty \frac{\lambda^i}{i!} 
\sum_{n=0}^i b(i-n;i,\mu)t_1^{i-n} t_{1+n}
\right)^{k-m} \nonumber  \\&
\times \left(
e^{-m\lambda}\sum_{i=0}^\infty \frac{ (m\lambda)^i}{i!} 
\sum_{n=0}^i b(i-n;i,\mu)t_1^{i-n} t_{j+m+n}
\right) .\label{eq:bigtau1}
\end{align}

We may now differentiate this to find expectations. Again we have to consider treating $t_1$ differently to $t_m$ for $m>1$.
\begin{align*}
D_1 \tau_j(\tvec{t})\vert_{t=1}
&=  e^{-j\lambda}\sum_{k=0}^\infty \frac{ (j\lambda)^k}{k!}  \sum_{m=0}^k b(k-m;k,\mu) \\&\bigg[
\left(
(k-m)e^{-\lambda}\sum_{i=0}^\infty \frac{\lambda^i}{i!} 
\sum_{n=0}^i b(i-n;i,\mu)( i-n + \delta_{n,0})
\right)    \\
&+
\left(
e^{-m\lambda}\sum_{i=0}^\infty \frac{ (m\lambda)^i}{i!} 
\sum_{n=0}^i b(i-n;i,\mu)( i-n + \delta_{1,j+m+n})
\right)
\bigg].
\end{align*}

The Kroeneker delta term $\delta_{1,j+m+n}$ can only be nonzero if both $j=1$ and $m=n=0$. Thus we treat $D_1 \tau_1$ and $D_1 \tau_j, j>1$ separately.
We shall use that
$\sum_{n=0}^i (i-n)b(i-n;i,\mu) = i\mu$ and that $b(i;i,\mu) = \mu^i$.

\begin{align*}
D_1 \tau_1(\tvec{t})\vert_{t=1}
&=
e^{-\lambda}\sum_{k=0}^\infty \frac{ (\lambda)^k}{k!}  \sum_{m=0}^k b(k-m;k,\mu)\\&\bigg[
\left(
(k-m)e^{-\lambda}\sum_{i=0}^\infty \frac{\lambda^i}{i!} (i\mu + \mu^i)
\right)+
\left(
e^{-m\lambda}\sum_{i=0}^\infty \frac{ (m\lambda)^i}{i!} 
(i\mu + \mu^i\delta_{m,0})
\right)
\bigg],
\end{align*}
which simplifies to
\begin{align*}
D_1 \tau_1(\tvec{t})\vert_{t=1}
&=
e^{-\lambda}\sum_{k=0}^\infty \frac{ (\lambda)^k}{k!}  \sum_{m=0}^k b(k-m;k,\mu)\bigg[
\left(
(k-m)(\lambda\mu + e^{-\lambda(1-\mu)}
\right)
+ 
\left(
m\lambda\mu + \delta_{m,0}
\right)
\bigg]\\
&=
e^{-\lambda}\sum_{k=0}^\infty \frac{ (\lambda)^k}{k!} 
\left(
k\mu(\lambda\mu + e^{-\lambda(1-\mu)}) + k\lambda\mu(1-\mu) + \mu^k
\right)\\
&=
\lambda\mu(\lambda\mu + e^{-\lambda(1-\mu)})
+ \lambda^2\mu(1-\mu) 
+ e^{-\lambda(1-\mu)}\\
&= \lambda^2\mu + (1 + \lambda\mu)e^{-\lambda(1-\mu)}.
\end{align*}

For $j>1$ we have
\begin{align*}
D_1 \tau_j(\tvec{t})\vert_{t=1}
&=
e^{-j\lambda}\sum_{k=0}^\infty \frac{ (j\lambda)^k}{k!}  
\sum_{m=0}^k b(k-m;k,\mu)
\\& 
\times \bigg[
\left(
(k-m)e^{-\lambda}\sum_{i=0}^\infty \frac{\lambda^i}{i!} (i\mu + \mu^i)
\right)
+
\left(
e^{-m\lambda}\sum_{i=0}^\infty \frac{ (m\lambda)^i}{i!} 
(i\mu )
\right)
\bigg]\\
&=
e^{-j\lambda}\sum_{k=0}^\infty \frac{ (j\lambda)^k}{k!}  
\sum_{m=0}^k b(k-m;k,\mu)
\times \bigg[
(k-m)(\lambda\mu + e^{-\lambda(1-\mu)}) + m\lambda\mu
\bigg]\\
&=
e^{-j\lambda}\sum_{k=0}^\infty \frac{ (j\lambda)^k}{k!} 
\left[
k\mu(\lambda\mu + e^{-\lambda(1-\mu)}) + k\lambda\mu(1-\mu)
\right]\\
&= j\lambda\mu(\lambda\mu + e^{-\lambda(1-\mu)})
+ j\lambda^2\mu (1-\mu)\\
&= j\lambda^2\mu + j\lambda\mu e^{-\lambda(1-\mu)}.
\end{align*}

We now look at $D_z \tau_j$ for $z > 1$. Differentiating \eqref{eq:bigtau1} 
we obtain
\begin{align*}
D_z\tau_j(\tvec{t}) \vert_{t=1}&=
 e^{-j\lambda}\sum_{k=0}^\infty \frac{ (j\lambda)^k}{k!}  
 \sum_{m=0}^k b(k-m;k,\mu)\\&
\times \bigg[
\left(
(k-m)e^{-\lambda}\sum_{i=0}^\infty \frac{\lambda^i}{i!} b(i-n;i,\mu)\delta _{z-1, n}
\right)\\
&+\left(
e^{-m\lambda}\sum_{i=0}^\infty \frac{ (m\lambda)^i}{i!} 
 b(i-n;i,\mu)\delta _{z-j-m,n}
\right)
\bigg]\\
&=
 e^{-j\lambda}\sum_{k=0}^\infty \frac{ (j\lambda)^k}{k!}  
 \sum_{m=0}^k b(k-m;k,\mu)\\&
 \times \bigg[
\left(
(k-m)e^{-\lambda}\sum_{i=0}^\infty \frac{\lambda^i}{i!}
\binom{i}{z-1}\mu^{i-(z-1)}(1-\mu)^{z-1}
\right)\\
&+
\left(
e^{-m\lambda}\sum_{i=0}^\infty \frac{ (m\lambda)^i}{i!} 
\binom{i}{z-j-m}\mu^{i-(z-j-m)}(1-\mu)^{z-j-m}
\right)
\bigg]\\
&=
 e^{-j\lambda}\sum_{k=0}^\infty \frac{ (j\lambda)^k}{k!}  
 \sum_{m=0}^k b(k-m;k,\mu)\\&
\times \bigg[
(k-m) \frac{(\lambda(1-\mu))^{z-1}}{(z-1)!}e^{-\lambda(1-\mu)}
+
 \frac{(m\lambda(1-\mu))^{z-j-m}}{(z-j-m)!}e^{-m\lambda(1-\mu)}
 \bigg]\\
 &=
 j\lambda\mu \frac{(\lambda(1-\mu))^{z-1}}{(z-1)!}e^{-\lambda(1-\mu)}
 +\\
 &
 e^{-j\lambda}\sum_{k=0}^\infty \frac{ (j\lambda)^k}{k!}  
 \sum_{m=0}^k b(k-m;k,\mu)  \frac{(m\lambda(1-\mu))^{z-j-m}}{(z-j-m)!}e^{-m\lambda(1-\mu)}.
\end{align*}
Note that if $m=0$ then only if $z=j$ will the term on the summation give a non-zero value, and if $z=j$ then this value will be $b(k;k,\mu) = \mu^k$. There is no obvious reduction of the $m$-summation, so it needs to be carried out explicitly. 
However we have the constraint that $z \ge j+m$ in order to get non-zero contributions from the inner term.

\subsubsection{Putting it all together}

In analogy to \eqref{eq:gen2f}, we have
\begin{equation}
G_3(\tvec{t}) = G(\tvec{\tau}(\tvec{t})), \label{eq:gen3c}
\end{equation}
from which we obtain
$$E[C_r^*] =  C^* \tilde{f}_r^* = C \sum_j \tilde{f}_j D_r \tau_j(\tvec{t})\bigg\vert_{ \tvec{t}=\tvec{1}},$$
where
\begin{align*}
D_1\tau_1\vert_{\tvec{t}-\tvec{1}} &= \lambda^2\mu + (1 + \lambda\mu)e^{-\lambda(1-\mu)},\\
D_1\tau_j\vert_{\tvec{t}-\tvec{1}} &= j\lambda^2\mu + j\lambda\mu e^{-\lambda(1-\mu)},
\end{align*}
and for $z>1$
\begin{align*}
D_z\tau_j \vert_{\tvec{t}-\tvec{1}}&=
 j\lambda\mu \frac{(\lambda(1-\mu))^{z-1}}{(z-1)!}e^{-\lambda(1-\mu)} +\\
 & e^{-j\lambda}\sum_{k=0}^\infty \frac{ (j\lambda)^k}{k!}  
 \sum_{m=0}^k b(k-m;k,\mu)  \frac{(m\lambda(1-\mu))^{z-j-m}}{(z-j-m)!}
 e^{-m\lambda(1-\mu)}.
\end{align*}
 
Note that the  multiplier of $b(k-m;k,\mu)$ is the probability
$P(X = z-j-m)$ where $X \sim \mbox{Poisson}(m\lambda(1-\mu))$.
Thus, evaluating these derivatives, substituting them into \eqref{eq:gen3c}
using $C=1$  and the $\tilde{f}_j$, and normalizing the
 $C^* \tilde{f}_r^*$ terms to sum to 1, we obtain the
haplotype cluster size probabilities $\tilde{f}_r^*$ for the three consecutive generations combined.

\subsection{Distribution of numbers of matching haplotypes}
The probabilities $\tilde{f}_k^*$ computed above are for 
proportions of  clusters  of equivalence classes $k$ identical haplotypes. Thus for example, $\tilde{f}_3^*$ is the probability that if a haplotype equivalence class was picked at random from the set of equivalence classes, then the equivalence class would consist of 3 identical haplotypes. 
What we want instead is: given a live male haplotype profile, the probability
$p_h(k)$ that there is a total of $k$ living males having that haplotype profile. These two probabilities are simply related:
$$
p_h(k) = \frac{ k \tilde{f}_k^*}{\sum_k  k \tilde{f}_k^*}.
$$
The expected number of living males having the same haplotype as a randomly selected living male is given by the expectation 
$$ \sum_k p_h(k) = \frac{\sum_k k^2\tilde{f}_k^*}{\sum_k k \tilde{f}_k^*}.$$
This indicates that the probabilities $\tilde{f}_k^*$ should be evaluated for a sufficiently high enough number of terms such that, when evaluating $ \sum_k kp_h(k))$, the high $k$ terms
$k^2 \tilde{f}_k^*$ have negligible contribution to the expectation.

\begin{figure}
\includegraphics[scale=0.6]{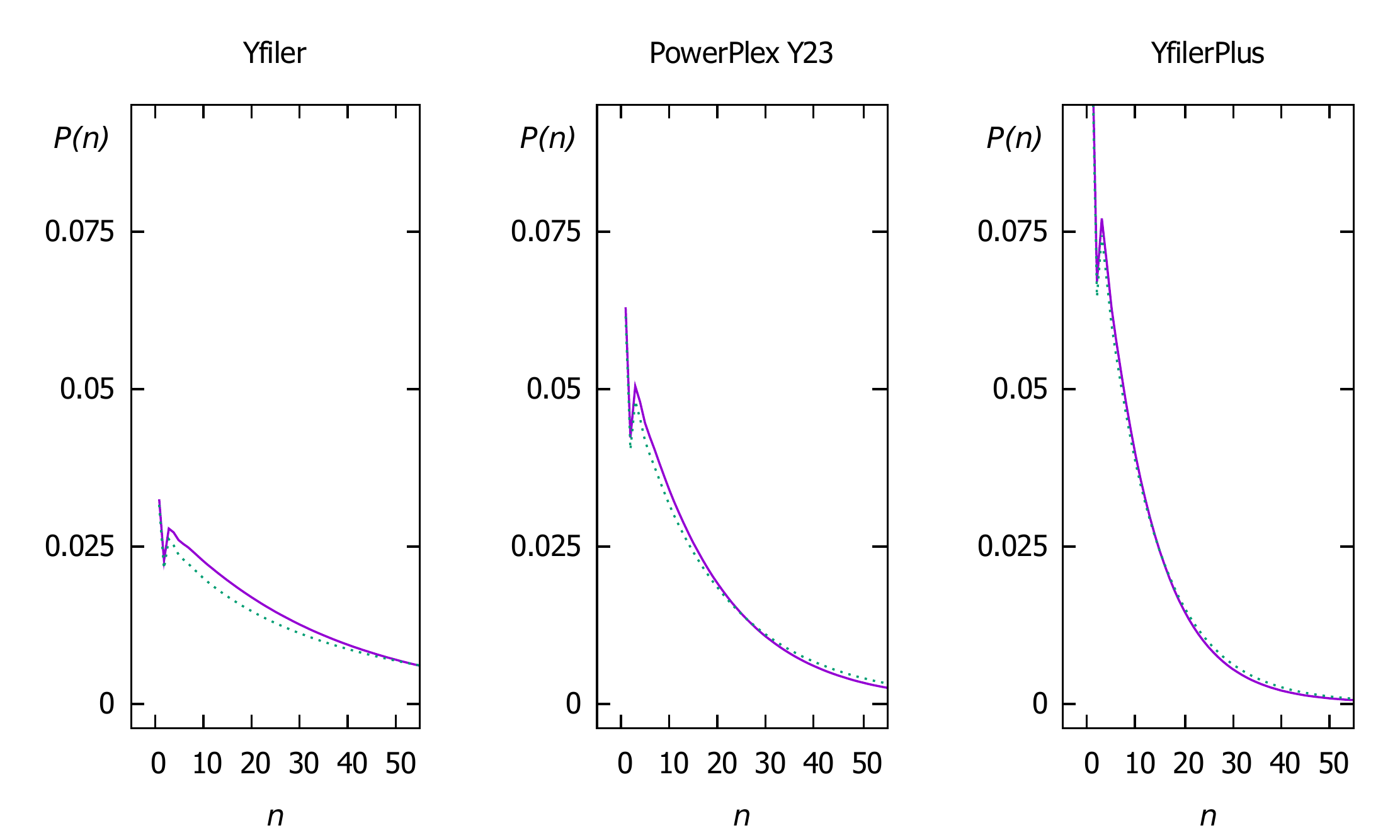}
\caption{Distributions of the number of live males spanning three generations having the haplotype profile of a given live male for three profiling kits, 
for constant population size (solid line)  and  for a growth rate of 2\% per generation (dotted line), derived from
 the multivariate probability generating function model in \secref{sec:subcriticalbp}.
}
\label{fig:clusters}
\end{figure}

In \figref{fig:clusters} we show the distributions obtained 
based on the mutation rates of loci in the  Yfiler, PowerPlex Y23 and YfilerPlus kits, for a  large constant population size, and for a large population with
a growth rate of  2\% per generation  $(\gamma=0.02)$; the plots are very similar to the Wright-Fisher plots in Figure~4 of
\citep{andersen2017convincing}.
For the plots of  \figref{fig:clusters}, 512 terms were retained and 200 iterations were used to estimate the $\tilde{f}_k$ probabilities.
The probabilities of the  $\tilde{f}_{512}$ values for each of the plot evaluations range from approximately $10^{-14}$  to $10^{-22}$.

Apart from the avoidance of using simulation to generate the distribution, with its associated time costs and estimation errors, another 
feature of the generating function model is that we do not require that our  haplotypes have alleles that are all integer repeats - we only require the mutation rates
on each locus and the population growth rate which is combined into the single Poisson rate $\lambda$ given above. Thus alleles with partial repeats can be  included within  the model, if we assume that such alleles  mutate at the same rate as integer repeat-valued alleles. In addition, haplotypes with deleted loci and loci with duplications are readily included, these features enter into the overall mutation rate. (Note that if a locus is deleted it remains deleted and thus has zero mutation rate.)
We now present a novel general framework for analysing Y- haplotype mixtures which incorporates these extensions: \secref{sec:framework} presents the  framework, and \secref{sec:match}
proposes a specific instance of the framework that  uses the haplotype cluster distribution model.

\section{A new framework for analysing Y-STR samples}
\label{sec:framework}

In the following we assume, for simplicity, that we have a single replicate amplified with a marker system consisting only of Y-STR loci, and that all persons, whether typed or not, are male. Let $\evid$ denote the peak height information on the set of loci
$\yloci$. Let $\kns$ denote the haplotype profiles of 
a set of individuals profiled on the loci $\yloci$, and $\ukns$ a non-empty set of untyped individuals, who are assumed under some hypothesis $H$ to be contributors to the sample-replicate in addition to some or all of the
profiled individuals.  
We shall assume that there is no \textit{known} familial relationship between any of the profiled individuals and the hypothesised untyped individuals.
Let $g_u$ be a possible specific joint profile of the $\ukns$. The problem at hand is to evaluate the likelihood
\begin{equation}
P( \evid \cd \kns, H) = \sum_{g_u} P(\evid \cd \kns, g_u, H) p(g_u\cd \kns),
\label{eq:like1}
\end{equation}
where we sum over all possible combinations of joint genotypes $g_u$ of the untyped persons $\ukns$ that are contributors under the hypothesis $H$.  
(Note that $p(g_u\cd \kns, H) = p(g_u\cd \kns)  $, because the profiles of the individuals is conditionally independent of  hypothesis $H$  specifying the contributors to the sample. In contrast, we do not assume that
$ p(g_u\cd \kns) =  p(g_u)$; the model presented in \secref{sec:match}
in concerned with evaluating $ p(g_u\cd \kns)$ taking into account the
observed profiles $\kns$.)

For simplicity we have omitted explicit mention of other conditioning parameters such as contributor DNA amounts and sample degradation
in \eqref{eq:like1}.  Given these other parameters, we may
factorize the peak-height evidence part into a product over the loci, thus

\begin{equation}
P( \evid \cd \kns, H) = \sum_{g_u} \left(\prod_{y \in \yloci}P(\evid_y \cd \kns, g_{u_y}, H) \right) p(g_u\cd \kns) , 
\label{eq:like2}
\end{equation}

where $\evid_y $ denotes the peak height information on the individual locus $y$, and 
$g_{u_y}$ denotes the joint profile of the $\ukns$ on the locus $y$.

If the $\yloci$ were autosomal, then we could factorize $p(g_u\cd \kns)$ into a product over the loci, say $p(g_u\cd \kns) = \prod_y p(g_{u_y}\cd \kns)$ which would mean the
$P( \evid \cd \kns, H)$ evaluations could be broken down into calculations for individual loci, and the results multiplied together, the so called \textit{product rule}. However this is inappropriate for Y haplotypes, and we have to consider the sum over the haplotypes $g_u$. 
Even for a single untyped male,  for marker systems of Y-STR loci used in laboratories the number of possible  terms in the summation is astronomical. 
For example, with a marker system consisting of 20 Y-STR loci, each with 10
alleles in the allelic ladder, the number of possible haplotypes for a single person would be 
$10^{20}$ even excluding the possibility of deletions or duplications. Allowing for deletions and duplications, there would be  66 possible profiles on each locus, leading to $66^{20} \approx 2.46 \times 10^{36}$ possible haplotypes over the 20 loci. 

The approach we take is to restrict this unmanageable sum over all possible haplotypes
to a much smaller and manageable sum over haplotypes that are deemed to be important. 
A version of this is proposed by \cite{taylor2018likelihood}, but requires having the profile  of a person of interest. 
Here we take a different approach to generating a reasonable subset of haplotypes, one in which the choice is driven by the information in the mixture, and can be used for mixture deconvolution even in the absence of a profiled person of interest. The computation may be broken down into the following sequence of steps, in which the values of $k$ and $m$ are inputs.

\subsection*{Steps in likelihood maximisation for Y haplotypes}
\begin{enumerate}
\item Maximise the likelihood (over contributor DNA amounts and sample degradation), \textit{initially}
 treating the Y-STR loci \textit{as if} independent by using the product rule. 
Population allele frequency data for each locus is used to inform the evaluation of the 
$p(g_{u_y})$ profile probabilities on the locus profiles $g_{u_y}$ on the locus $y$.
\item Conditional on the maximum likelihood estimates, find for each locus $y \in \yloci$ the set of (up to) $k$ joint profiles that  are in the $k$ highest  likelihood terms contributing to the overall likelihood 
$\sum_{u_y}P(\evid_y \cd \kns,g_{u_y})p(g_{u_y}\cd \kns)$ on each locus separately.
\item Combine these single locus  profiles to produce joint haplotype profiles over all the loci that have (up to) the  $m$ highest likelihoods, (highest with respect to the product rule), denote these joint haplotypes by $\hap$.
\item For each $h \in \hap$ calculate the profile probability $p(h\cd \kns, \dbase)$, where
$\dbase$ is a database of observed haplotypes.
\item 
Re-maximise the likelihood (with respect to contributor amounts and sample degradation parameters), this time over 
$ \sum_{h \in \hap} P(\evid \cd h, \kns)p(h\cd \kns, \dbase)$ 
and approximate $$P( \evid \cd \kns) \approx \sum_{h \in \hap} P(\evid \cd h, \kns)p(h\cd \kns, \dbase).$$
\end{enumerate}

Step 1 is standard likelihood maximization such as carried out for example in Euroformix \citep{bleka2016euroformix} and the DNAMixtures package \citep{graversen2014statistical}. (At present, these packages
are restricted to amplification kits having only autosomal loci and possibly Amelogenin. However both are open source and could be adapted to be used for Y-STR loci.)

Step 2 is carried out using highly efficient algorithms developed for Bayesian networks
\citep{nilsson:97,cowell:book}. We say ``(up to) $k$" because there may be fewer than $k$ possible joint-profiles on a locus. 

Step 3 is relatively simple to implement and can be carried out efficiently, even though there  are possibly up to $k^{\vert \yloci\vert}$ combinations to consider. Essentially, one keeps the $\vert \yloci\vert$ sets of single loci profile terms in separate arrays or stacks, each sorted from highest to lowest likelihood. A breadth first search can then be implemented (for example) to find the top $m$
highest likelihood haplotypes 
(very similar to the algorithm used in Step~2). Note for example the highest joint locus profile will be the combination of each of the  highest single locus profiles. The second-highest likelihood haplotype will differ in only one locus, and at the locus the profile will be the second term on the ordered stack.
(Thus it is not necessary to  first generate all the possible combinations.)

Step 4, in essence, `strips out' the product-rule profile probabilities from the haplotype likelihoods, 
to be replaced by haplotype probabilities. A specific model  for this step is proposed  in \secref{sec:match}. 
 
Step 5 does a final likelihood maximization of parameters using this restricted set of haplotypes.
Note that  the profile probability estimates from Step 4 do not change for Step 5, because we do not change the set 
of haplotypes $\hap$ used in the likelihood maximization. In addition, because the haplotypes $h$ are  fully defined on all loci, their peak-height likelihoods are readily evaluated. (In the example presented in \secref{sec:realdata},  re-maximization over contributor amounts only was carried out, because the degradation parameter estimated from step 1 was equal to zero.)
 
Note that the steps above provide a general algorithm, which is specialized only by the choice of 
 haplotype probability model used in Step 4.
 
The rationale of the algorithm is that, while the product rule is deficient in regards to 
haplotype profile estimation, nevertheless the highest likelihood profiles
found on each locus will have a high likelihood mainly because of their explanatory power with regard to the observed (and also  unobserved) allelic peak heights, rather than from the genetic profile probabilities of the untyped contributors.  Thus, combining across loci 
these high likelihood profiles,  we form a candidate set of haplotypes that are expected between them to  have a high explanatory power with regard to the peak height data.
 These haplotypes need to be re-weighted,   by replacing the deficient product rule probabilities with more realistic haplotype profile probabilities, for the final overall maximum likelihood estimate. 

Note that after the final maximization, the individual haplotypes can be sorted by likelihood value. The difference between the  highest and lowest haplotype likelihoods is then readily evaluated, which can give an indication as to whether or not the  numbers $k$ and $m$ used  in Steps 2 and 3 are large enough.

\section{A Y haplotype match probability model}
\label{sec:match}

Now let us recap on what we are trying to achieve. We have a set of joint haplotypes $\hap$ over untyped males,  generated  in Step ~3 of the algorithm in \secref{sec:framework}. Each of the individual untyped person's haplotype $h_u, u \in U$,  of a joint haplotype $h\in \hap$ may have locus profiles with 
deletions, duplications, and  partial repeats that may or may not match individual haplotypes in a population database $\cal D$ of complete profiles and/or  any haplotype profiles of fully typed persons. We want to find a match probability taking this information into account.

We first make the assumption that the probability for the joint haplotype is the product over the probabilities of each individual untyped person's haplotype, that is $P(h) = \prod_u P(h_u)$. 
To proceed further, we introduce the notion of \textit{haplotype patterns}, for which we have ordered the loci in the haplotype by their position in the Y-chromosome. 
Suppose that the number of loci under consideration is $N$.
We define three different haplotype patterns for a given profile:
\begin{enumerate}
\item Identity pattern $I$: This is an ordered list of the $N$ locus profiles.
\item Deletion/duplication pattern $D$: This is an ordered list of $N$ zeros, ones and twos, in which we have  0 if the locus is deleted, 1 if it is not and occurs singly, and
2 if the locus occurs duplicated. (Although triplications do occur, albeit very rarely, they will be ignored in our analysis.)
\item Repeat pattern $R$: This is an ordered list of $N$ pairs of integers in which the integer denotes the repeat parts of the alleles on the locus.
If the locus is deleted, it has the pattern $(0,0)$. If the locus occurs singly the pair is $(0,r)$ where $r$ is the repeat part of the allele. If the locus is duplicated
the pair is $(r_1, r_2)$ where $r_1$ and $r_2$ are the repeat parts of the observed alleles, and we order the pair so that  $r_1 \le r_2$.
\end{enumerate}

A comment is in order concerning mutations and the  $D$ and $R$ patterns.
We assume that alleles with partial repeats mutate to other alleles having the same partial repeat number (for example 12.2 could mutate to 11.2 or 13.2),
so that changes to the repeat part do not occur, for example 12.2 mutating to 12 or 12.3 is not allowed. 
Although the latter types of mutation do occur in practice, their rate of occurrence is much lower than that of the repeat-preserving mutations that we can, to a good approximation, ignore them. We also assume that the mutation rate of a locus does not depend on the
value of the repeat part. (This means that the  alleles of a locus with repeat parts
mutate at the same rate as alleles having an integer number of repeats.)
 We also assume that the rates of locus deletions or duplications per generation is so low that both can be ignored (although historically they have occurred). Under these assumptions, we will have that the offspring of a  haplotype having specific  $D$ and $R$ patterns will all have  $D$ and $R$ patterns identical to the parent haplotype, although specific alleles may differ due to mutation. This means that the  branching process model introduced earlier will be applicable to haplotypes regardless of the presence or absence of deletions, duplications of partial repeats.

Using the above patterns and assumptions, a high-level description for estimating $P(h_u)$ is as follows, in which
 $M$ is the sum of  number of males in the population database and  the number of typed males, (i.e., $M =  \vert {\cal D} \vert + \vert \kns \vert$),
and $\Omega$ is the total number of males in the population. 

\begin{enumerate}
\item Find the Identity pattern $I_u$ of $h_u$, and the distribution (not conditioned on $M$) $P(N\cd \Omega)$ of matching profiles $N$. 
Note that $P(N\cd \Omega)$ will depend on the deletion/duplication pattern $D$, because this affects the overall non-mutation rate.
For example, if a locus is deleted, it cannot mutate. Alternatively,  if a locus is duplicated
its \textit{non-mutation} probability will be of the form $(1-\mu)^2$ instead of
$(1-\mu)$.
\item Count the number $c_I$ of  haplotypes  amongst the population database $D$ and typed persons $\kns$ that match $h_u$. (Note that
If $M=0$, that is, if there is neither a population database $\cal D$ of reference haplotypes nor a set of 
typed persons $\kns$, then necessarily we have $c_I = 0$ .)

\item Compute the distribution
$P(N \vert c_I, M, \Omega)$, and its expectation 
$$E[N\vert c_I, M, \Omega] = \sum_n n P(n \vert c_I, M, \Omega).$$

and define $$ p_u =  \frac{E[N\vert c_I, M, \Omega]}{\Omega}.$$

\item 
 If $c_I >0$ then  set   $P(h_u) = p_u$.
 \item Otherwise, if $c_I=0$,  we continue as follows:
\begin{enumerate}
\item  Find the deletion/duplication pattern $D_u$ of the  untyped male, and count the 
 number $c_D$ of matching deletion/duplication patterns in the haplotypes amongst the population database ${\cal D}$ and typed persons $\kns$. 

\item Find the repeat pattern $R_u$ of the untyped male, and count the 
 number $c_R$ of matching  repeat patterns in the haplotypes amongst the population database $D$ and typed males $\kns$. 
\item Make a multiplicative adjustment of $p_u$ based on the patterns $D_u$, $R_u$ and counts $c_D$, $c_R$, and set this adjusted value to 
  $P(h_u)$.
\end{enumerate}
\end{enumerate}

The distribution $P(N\cd \Omega)$ of Step~1 corresponds what is plotted in \figref{fig:clusters}, and for large $\Omega$ will have little dependence on the
population size as shown in the simulations of \citep{andersen2017convincing}.

The  distribution $P(N \vert c_I, M,\Omega )$ of Step~3 is computed in \citep{andersen2017convincing} by simulation  using importance sampling, assuming that the database profiles  are a random sample from the population. If we make the same database assumption, 
then the distribution $P(N\vert c_I, M, \Omega)$ may be computed as follows. 

Let $P_{hyper}(k\vert n, m, \Omega)$ denote the hypergeometric distribution probability, for a population of size $\Omega$
of which there are $n$ objects of the first  kind and $\Omega -n$ objects of the second kind, that if $m$ objects are randomly selected  without replacement from the population then $k$ of the $m$ objects will be of the first kind. 

Let $C_I$ denote the random variable for the number of identity pattern matches, of which $c_I$ were actually seen, given that there are a total of $n$
such matching haplotype identity patterns in the population.
Then we set
\begin{equation}
P(C_I\vert n, M, \Omega) = P_{hyper}(C_I+1,n,M+1, \Omega)/(1-P_{hyper}(0,n,M+1, \Omega)).
\label{eq:condhg}
\end{equation}
In \eqref{eq:condhg} we treat the haplotype $h_u$ under consideration
as having been seen, which has the effect of increasing the observed count by 1, hence the use of the sample size $M+1$ in the hypergeometric distributions. The $C_I$ denotes the number of matches to $h_u$, but it does not include  the haplotype $h_u$, hence we have `$C_I+1$' in the numerator term to include the haplotype $h_u$ of $u$ in the total count. In turn, this requires conditioning the hypergeometric distribution in the numerator
on $h_u$ of $u$ definitely being sampled; the conditioning is effected by the denominator in 
\eqref{eq:condhg}.

Bayes' theorem may now be applied to find
\begin{align*}
P(N=n \vert c_I, M, \Omega) 
&\propto  P(C_I = c_I\vert N=n, M, \Omega)P(N=n\cd \Omega)\\
&=  \frac{P(C_I = c_I\vert N=n, M, \Omega)P(N=n\cd \Omega)}{P(C_I=c_I \cd M, \Omega)},
\end{align*}
where
$$
P(C_I=c_I \cd M, \Omega) = \sum_n P(C_I = c_I\vert N=n, M, \Omega)P(N=n\cd \Omega),
$$
from which $p_u$ as defined in Step~3 above may be found.

If $c_I >0$, then we are done by setting $P(h_u) = p_u$, as in  Step~4. Otherwise we make multiplicative adjustments to $p_u$ based on the
deletions, duplications, and the repeat patterns (Step~5) as  elaborated in the next section. 
 
\subsection{Multiplicative adjustments for deletion and duplication patterns}
\label{sec:dupdels}
There are two cases to consider. 

\subsubsection{$c_D >0$}
If $c_D >0$, then the adjustment factor is set to 

$$f_D = \frac{1 + c_D}{M+1}$$
 which is the relative proportion of observed haplotypes having matching deletion and duplication patterns amongst the profiles in the database and typed persons; we add 1 in the numerator and denominator to also include the haplotype of the untyped person $u$ under consideration.

\subsubsection{$c_D =0$}
On the other hand, if $c_D =0$, then we find   factors separately for the deletions and the duplications, and combine them multiplicatively. 
Note that a locus cannot both be duplicated and deleted.
We first look at deletions, introducing two parameters $a$ and $b$.
 To evaluate the deletion factor we examine the loci in  order (along the chromosome).
If the first locus is a deletion, we have a factor $f = a$, otherwise we have a factor $f=1$. We now go through the remaining loci in sequence. 
If a locus is deleted, but the previous locus is not, then we update $f$ by a factor $a$; thus $f := f*a$. If the locus is deleted and the previous one is also deleted,
 then we update the factor  $f$ by the factor $b$; thus $f := f*b$. Otherwise if the locus is not deleted we leave $f$ as it is.
Essentially, the factor $a$ captures the rarity of deletion in a population, the factor $b$  models correlations of  deletions 
in  loci that are close together in the chromosome; it gives a geometric weight to the number of consecutive deletions. Thus for example a sequence of $j$ consecutive deletions will have
a weight factor $f = ab^{j-1}$.  \tabref{tab:deleg} shows some examples to illustrate the calculation.

We now consider the duplication factors. This is similar to the calculation of deletion factors. 
We introduce to parameters $c$ and $d$ that represent factors of an initial duplication and for consecutive duplications. However, we need to take account that some loci, such as DYS385a/b, are always multicopy loci. We therefore do not include factors for such loci.
As for the deletions, we start at the first locus, and give a factor $g=1$ if the locus has deletion/duplication pattern value of 0 or 1,  or the locus is multicopy, otherwise we give a factor $c$. We then move along the loci in turn. If a locus has pattern 0 or 1, or has pattern 2 and is multicopy, then we leave $g$ unchanged. Otherwise 
the locus is single-copy but has a duplication. If the previous locus is also single-copy but has a duplication we update $g$ by the factor $d$, otherwise we are at the start of a new sequence of excess duplications and we update $g$ by a factor $c$ instead. \tabref{tab:deleg} shows some examples to illustrate the calculation.

The overall deletion/duplication factor is therefore
\begin{align}
f_D &= \frac{1 + c_D}{M+1}& \mbox{ if } c_D >0, \label{eq:fd1}\\
f_D &= f \times g &\mbox{ if } c_D =0.\label{eq:fd2}
\end{align}

\clearpage
\begin{landscape}
\begin{table}[htbp]
\caption{Illustration of deletion $(f)$ and duplication $(g)$ factors for three haplotype patterns.
In the first haplotype pattern $h_1$ there are no deletions, hence the deletion factor is $f=1$. There are no excess duplications, because
the locus DYS385a/b is a multicopy locus it does not lead to a factor $c$, hence we have $g=1$.
In the second haplotype pattern $h_2$  there are three groups of consecutive deleted loci. The one at DYS389-I is isolated so contributes a factor $a$. 
Then there are three consecutive deletions, which gives a factor of $ab^2$. There is a final isolated deletion having a factor $a$. 
Hence the overall deletion factor is $f = a^3b^2$. There is one isolated duplication, on DYS19, hence the duplication factor is $g=c$.
In the final haplotype pattern $h_3$ , there are two groups of deleted loci, giving a factor $f = a \times ab = a^2b$. There are also two groups of excess duplications
giving a factor of $g = cd \times c = c^2d$.
}
\begin{center}
{\small
\begin{tabular}{c|ccccccccc|c|c}
Pattern &DYS19 &DYS385a/b &DYS389-I &DYS389-II &DYS390 &DYS391 &DYS392 &DYS393 &DYS437 & $f$ & $g$ \\
$h_1$ &1 &2 &1 &1 &1 &1 &1 &1 &1  & 1 & 1\\
$h_2$ & 2 &2 &0 &1 &0 &0 &0 &1 &0  & $a^3b^2$ & $c$\\
$h_3$ &1 &2 &0 &2 &2 &0 &0 &1 &2 & $a^2b$ & $c^2d$\\
\end{tabular}
}
\end{center}
\label{tab:deleg}
\end{table}
\end{landscape}
\clearpage

\subsection{Multiplicative adjustments for repeat patterns}
\label{sec:repeatpats}
The treatment for repeat patterns is similar to that of the deletion and duplications, but we do not have a correlation between the repeat patterns of neighbouring loci, as they can be considered as arising from independent mutation events. 
Again we have two cases. 

\subsubsection{$c_R >0$}

If $c_R >0$, then we use an adjustment factor is given by 

$$f_R = \frac{1 + c_R}{M+1}.$$ 

\subsubsection{$c_R =0$}
On the other hand, if 
 $c_R =0$ we evaluate for each locus $m$ the number of times $r_m$ in the database and typed persons that the repeat pattern occurs.
 We then set the overall repeat factor to be
 $$f_R = \prod_m \frac{1 + r_m}{M+1}$$
In this way our factor captures the rarity, or otherwise, of the particular types of repeats on each locus.

The overall repeat pattern factor is therefore
\begin{align}
f_R &= \frac{1 + c_R}{M+1} &\mbox{ if } c_R >0, \label{eq:fr1}\\
f_R &= \prod_m \frac{1 + r_m}{M+1} &\mbox{ if } c_R =0.\label{eq:fr2}
\end{align}

\subsection{Overall repeat-pattern factors}

Taken together with the deletion/duplication factors, the overall value set for the haplotype probability $P(h_u)$ is given by
\begin{align*}
P(h_u) &= f_D f_R \,p_u\\
&=   f_D f_R \frac{E[N\vert c_I, M, \Omega]}{\Omega}.
\end{align*}

\subsection{Rationale of factors}
\label{sec:rationale}

The rationale for these pattern factors (applied when $c_I=0$) is that, if a haplotype $h_u$ is neither observed in a database nor in a 
set of typed individuals in a case, but the
pattern is, then conceivably the haplotype $h_u$ could have a common ancestor with an observed haplotype with which it
 shares the pattern, the haplotypes being  different due to mutations along the meiosis lines from the common ancestor. (Recall that under the approximations assumed about mutations, that deletion, duplication and repeat patterns are preserved under meioses.) The more haplotypes with matching patterns there are, the more possibilities there are for this to be the case, so the fraction that do match seems a reasonable factor with which to scale the probability $p_u$.   This explains of the factor choices in \eqref{eq:fd1} and \eqref{eq:fr1}: essentially they are estimates of the proportions of such patterns in the population.
 
On the other hand if the pattern on the haplotype
$h_u$ does not match any observed pattern, rather than assign the haplotype a probability of zero, the factors are heuristics that  down-weight the haplotype probability $P(h_u)$, so that the more artefacts the pattern of $h_u$ has the more  unlikely the 
haplotype is to occur.  This explains of the factor choices in \eqref{eq:fd2} and \eqref{eq:fr2}.

An important practical  consideration is specifying the  numerical factors $a, b,c$ and $d$. 
Estimates for these can be found using the large Y-haplotype database given in \citep{purps2014global}. 
The database consist of approximately 19,600 haplotypes. Of these 95 have deletions present, some with multiple isolated deletions. Ten have a pair of consecutive deletions, and there is one sequence of three consecutive deletions.  Hence a crude estimate of
$a$ is $\hat{a} = 95/19600 = 0.0048$, and of $b$ is $ \hat{b} = 10/95 \approx 0.105$.
The parameters for duplication may be found similarly. There are 126 haplotypes with pair duplicates, so we can make the estimate
$ \hat{c} = 126/19600 = 0.0064$. There are 8 haplotypes with two consecutive duplications, so an estimate for $d$ is 
$\hat{d} = 8/126 = 0.063$. 

Note that these estimates do not take into account that some regions of the Y chromosome are more prone to deletions than others, and similarly for duplications. In addition they have been estimated using all the populations in the dataset. A refinement would therefore be to have population and chromosome region specific parameters, (and a modification of the adjustment formulae given above),
however this would require a very much larger dataset than that  published in \citep{purps2014global}. 
However, one could argue that the biological processes driving such random deletions and duplications from generation to generation, take place within the cells of individuals, and so should not depend on the population. 
We shall return the matters of robustness and sensitivity
to these parameters in \secref{sec:robust}.

The reader may wonder why, when $c_I=0$ for a haplotype $h_u$, 
we do not simply set $p_u = E[N\cd 0,M,\Omega]/\Omega$ rather than using the scaling factors described in \secref{sec:dupdels} and \secref{sec:repeatpats}. To answer this, consider the following comparison of two unobserved haplotype profiles. 

The first haplotype profile $h_1$ over 20 single-copy Y-STR loci has no deletions or duplications, and the repeat parts of all its alleles are 0 (so that the alleles all have integer designations). Let the overall mutation rate (over all loci) for this haplotype be  $\mu$.
Now consider another   haplotype $h_2$ arising in the likelihood function, for which the profile is such that half of the loci are marked as deleted, and the profiles for the remaining 10 loci agree with those of $h_1$ but with alleles that are duplicated in all of these loci, such that the overall mutation rate is again $\mu$. 
Alternatively, let  
of each locus of $h_2$ have the same profile of  $h_1$ but with all alleles
having a non-zero added (the overall mutation rate will then be unaltered).
Under these circumstances, both haplotypes $h_1$ and $h_2$ would be assigned the same 
probability if the formula $p_u = E[N\cd 0,M,\Omega]/\Omega$ were to be used. However the number of deletions or duplications observed in real haplotypes is typically none or quite small, as shown in the figures cited above, and similarly for the number of loci having non-zero repeat parts. To give equal probabilities to $h_1$ and $h_2$ is thus not appropriate. Instead  
adjustment to the probabilities of unobserved haplotypes depending on their 
patterns of deletions, duplications and non-zero repeat parts, with 
multiplicative weights becoming smaller the more such artefacts there are, would seem more appropriate. The weighting factors described above provide one simple and intuitive \textit{suggestion} for doing this.

\section{An application to real sample data}
\label{sec:realdata}

We shall look at an example from a dataset produced and made publicly available by Boston University \citep{cotton:etal:2012}. This dataset consists of  around  2900 (fsa format) files  of laboratory prepared single source and mixed DNA samples amplified  with four kits, including, of interest  for this paper, one from the AmpFlSTR\regd\ Yfiler\regd\ kit.
 The samples were prepared by individually extracting DNA from  blood from   four persons, of which three were male and  one female. Mixtures were made by diluting and combining these single person extracts in various proportions post-extraction.

We shall analyse this data using software developed by the author, based upon the
probabilistic genotyping framework described in  \citep{cowell2018unifying}.
 This models the whole PCR process from sample to electropherogram, so the
details below about sample and replicate volume, etc., are not superfluous to the  analysis. Other details such as injection time and voltage are also relevant but have omitted for brevity. 
However, the laboratory setup is \underline{not} the DNA model that is implemented in the author's software, and not all of the information required for the software is available
(for example the DNA contributors were anonymous  of unknown population).
We shall therefore  analyse the data
\textit{as if} it were prepared under the following assumptions.

\begin{enumerate}
\item  We take as our Y-STR database the subset of European-American profiles publicly available from \citep{purps2014global}. 
\item All contributors, typed and untyped,  are assumed to be from the European-American population, with an assumed population size of two hundred million.
\item The sample volume of extracted DNA is 50 $\mu$L.
\item The extraction efficiency of DNA is 20\%.
\item 5 $\mu$L of the sample extract is used in the replicate, which is amplified in a total volume of 25$\mu$L.
\end{enumerate}

In essence, a Poisson distribution would be a reasonable model of the extraction and sampling of genomic strands from the cellular DNA 
prior to amplification, and we are  approximating the Poisson by a specific  binomial sampling model. 

\subsection{Male contributor profiles}

The male individuals are denoted by the letters A, C and D; their profiles are shown in \tabref{tab:profiles}. Note that on 4 of the 16 loci every person has the same profile.

\begin{table}[htbp]
\caption{Genetic profiles of the three individuals A,C and D, on the  loci of the Yfiler kit.}
\begin{center}
\begin{tabular}{l|ccc}
Locus &A &C &D \\ \hline
DYS19 &14  &15  &15  \\
DYS385-a-b &13/18 &16/17 &15/17 \\
DYS389-I &13  &14  &14  \\
DYS389-II &32  &31  &31  \\
DYS390 &23  &21  &21  \\
DYS391 &10  &10  &10  \\
DYS392 &11  &11  &11  \\
DYS393 &12  &14  &13  \\
DYS437 &14  &14  &14  \\
DYS438 &10  &11  &11  \\
DYS439 &11  &13  &11  \\
DYS448 &20  &21  &21  \\
DYS456 &15  &15  &15  \\
DYS458 &17.2  &17  &16  \\
DYS635 &20  &21  &21  \\
Y-GATA-H4 &11  &12  &10  \\ \hline
\end{tabular}
\end{center}
\label{tab:profiles}
\end{table}

\subsection{Replicate profile}
The particular mixture we shall look at was obtained from the Yfiler file \newline
\texttt{Y\_3\_SACD\_NG0.4\_R3,1,2\_A1\_V1.fsa}. The file naming convention means that
this is a three person mixture with contributors A, C and D, prepared with relative amounts of DNA  in the proportions 3:1:2 respectively. An estimated total of 0.4ng amplifiable DNA was amplified in the replicate with an injection time of 5 seconds (the time is coded by V1 at the end of the filename). Using the free open source Osiris software \citep{osiris} the author obtained from the fsa file the peak height values shown in \tabref{tab:peaks}. (In extracting the peaks using Osiris, a threshold of 5 RFUs was used.)
 The smallest peak height value retained is 15 RFUs, which is the value we use for the  analytic threshold in the analyses to follow.
 At this analytic threshold, all main alleles are present, and all other alleles are in stutter positions of these main alleles, with the exception of allele 11.1 of Y-GATA-H4, with a peak height of 27 RFUs. This allele is not present in any of the contributors, and might be an artefact coming from the peak of allele 11, or could be a drop-in artefact or noise. with a significantly higher setting of the analytic threshold, there would be some dropout.

\begin{table}[htbp]
\caption{Allelic peak heights for the Yfiler 3-person  mixture}
\begin{center}
\begin{tabular}{l|cl|cl|cl|cl}
Locus & Allele&height & Allele&height & Allele&height &  Allele&height  \\ \hline
DYS19 & 13&16 & 14&161 & 15&167  \\
DYS385-a-b & 12&25 & 13&258 & 15&63 & 16&72 \\ 
DYS385-a-b (cont)& 17&122 & 18&136  \\ 
DYS389-I & 12&24 & 13&267 & 14&235  \\
DYS389-II & 30&23 & 31&130 & 32&99 \\
DYS390 & 20&15 & 21&174 & 22&21 & 23&168  \\
DYS391 & 10&717 & 9&46  \\
DYS392 & 10&47 & 11&500  \\
DYS393 & 11&24 & 12&227 & 13&165 & 14&50  \\
DYS437 & 13&35 & 14&485  \\
DYS438 & 10&318 & 11&250  \\
DYS439 & 10&19 & 11&276   &13&23\\
DYS448 & 20&326 & 21&206  \\
DYS456 & 14&64 & 15&523  \\
DYS458 & 15&21 & 16&152 &16.2&38 & 17&90 \\
DYS458 (cont)  & 17.2&372  \\ 
DYS635 & 19&18 & 20&244 & 21&147  \\
Y-GATA-H4 & 10&108 & 11&106 & 11.1&27 & 12&67  \\ \hline
\end{tabular}
\end{center}
\label{tab:peaks}
\end{table}

\subsection{Maximum likelihood estimates}

In finding maximum likelihood estimates, modelling of stutters were included but forward and double-reverse stutters were excluded; it was found that this combination gave the highest likelihood for the first scenario, and so was used for the remaining scenarios. We used the estimates of the deletion and duplication factors given in \secref{sec:match}.

In \tabref{tab:results}
are shown the results of likelihood-maximization for the scenario (or hypothesis) that A, C and D are the contributors to the mixture (top row) and the 7 alternatives where we replace one or more contributors by an untyped male contributor. For each analysis an upper bound of 5,000 (joint) haplotypes was given for the search procedure. For the scenario  of three unknown males, the upper limit of the number of haplotypes to search over was also raised to half a million - this is shown in the final row of the table.

The second column of  \tabref{tab:results} gives the maximum likelihood estimates based on using the product-rule for the loci (Step 1 of the procedure of \secref{sec:framework}).
We see that on increasing the number of untyped persons substituted for typed persons, the maximized likelihood estimate decreases as expected. In column 3 of \tabref{tab:results} are the corresponding results using the haplotype probability model of this paper. This too follows the pattern of column 2 as more untyped persons are substituted, but now the likelihoods are decreasing at a  slower rate, with a difference of around 20 (on the $\log_{10}$ scale) when compared to the product-rule values in the  final two rows.

For a given defendant  $K$, a prosecution hypothesis for this data could be $K,U1,U2$; that is, that $K$ and two untyped males $U1$ and $U2$, all mutually of unknown relatedness, are the contributors to the DNA sample.
The defence hypothesis  $U1,U2,U3$ is that three untyped males, for which no known relatedness is known, are the contributors to the  DNA sample.
Log-likelihood ratios of such hypothesis pairs are shown in \tabref{tab:lratios}, in which
 each true contributor is treated as a defendant or  person of interest. In each of the three hypotheses comparisons, the values are much lower for the haplotype model compared to 
 those computed using the product rule.

The final three columns \tabref{tab:results} of give the estimated number of cells (in the sample prior to DNA extraction) under each scenario for the three contributors on the first column, in the respective orders. 
The estimated cell counts can be considered fictional because of the 
caveats given about the modelling of the sample preparation in \secref{sec:realdata}.
However,
the relative proportions of the estimated counts does provide a way to check concordance with the data. The 
experimentally prepared ratios of DNA amounts were 3:1:2.
In relative amounts, to the overall total, this would be for each contributor 
$0.500 : 0.167 : 0.333$, which is broadly in line with the values in
\tabref{tab:results}. Finding the relative amounts in each row, and taken the average relative amount for each contributor, we obtain
$0.533 : 0.155 : 0.311$.

A  concordance check with the 
sample preparation can also be found using the estimated counts combined with the parameters of the modelling as follows. The total sum of the estimated cell counts is the amount of DNA pre-extraction and sampling for the replicate prior to amplification. Now the average total number of  estimated cells in each scenario in \tabref{tab:results} is 3095.75 (excluding the final row). 
Using the assumed extraction efficiency of 0.2, and the assumed $5/50 = 0.1$ fraction of the sample put into the replicate for amplification, the estimated amount of amplifiable DNA in the replicate will be  $3095.75 * 0.2 * 0.1 \approx 61.95$ cells.
This equates to approximately $61.35*6.6 \approx 408.7$pg of amplifiable DNA, in line with the 400pg of DNA reported for the mixture.

\begin{table}[htbp]
\caption{Maximized log-likelihood estimates of eight scenarios for the Yfiler mixture; logarithms are to base10. The ordering of the cell count estimates for each scenario in the final column follows that of the three hypothesised contributors in the first three columns. In all scenarios except for the final line of the table, an upper bound value of $m = 5000$ joint haplotypes was set in the search procedure. For the final row, the limit was set to $m = 5000000$. All scenarios used $k=1500$.}
\begin{center}
\begin{tabular}{ccccccc}
Scenario &Log-likelihood & Log-likelihood  &\multicolumn{3}{c}{Cell number} \\
 & (product rule) & (Haplotype model) &\multicolumn{3}{c}{ estimates} \\ \hline
A C D &-101.045 &-101.045 &1737 &453 &903 \\
A C U &-114.077 &-106.609 &1673 &550 &876 \\
A U D &-112.368 &-105.032 &1651 &465 &976 \\
U C D &-115.445 &-106.892 &1737 &477 &903 \\
A U U &-124.405 &-109.540 &1578 &438 &1071 \\
U C U &-128.388 &-112.663 &1675 &550 &871 \\
U U D &-126.768 &-111.089 &1653 &464 &975 \\
U U U &-135.648 &-116.370 &1516 &448 &1126 \\ \hline
U U U &-135.649 &-115.481 &1559 &442 &1089 \\ \hline
\end{tabular}
\end{center}
\label{tab:results}
\end{table}

\begin{table}[htbp]
\caption{Log-likelihood ratios for a sample of hypothesis pairs, using product rule and 
haplotype model maximized likelihoods; logarithms are to base 10. For the scenario with three untyped males, 
the values from the last row of \tabref{tab:results} based on 500,000 joint haplotypes was used.}
\begin{center}
\begin{tabular}{ccc}
Hypotheses & log likehood ratio &log likehood ratio \\
& product rule & haplotpe model \\ \hline
AUU vs UUU &11.244 &5.941 \\
CUU vs UUU &7.261 &2.818 \\
DUU vs UUU &8.881 &4.392 \\ \hline
\end{tabular}
\end{center}
\label{tab:lratios}
\end{table}

\begin{figure}[htbp]
\begin{center}
\includegraphics[scale=0.6]{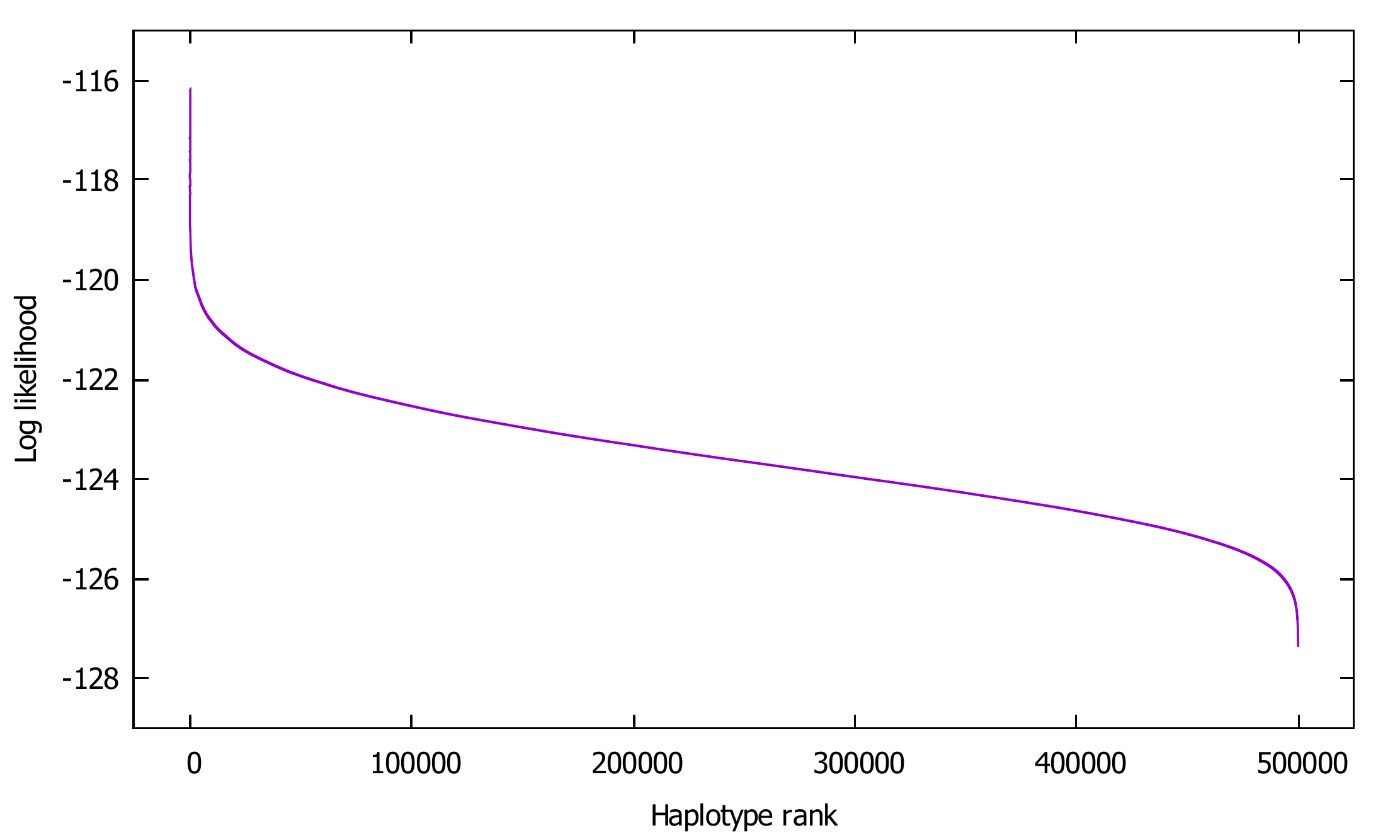}
\end{center}
\caption{Ordered log-likelihood values (base 10) of the 500,000 haplotypes generated for the three untyped males scenario.
\label{fig:hapdist}}
\end{figure}

In \figref{fig:hapdist} is a plot of the ordered log-likelihood values of each of the joint-haplotypes generated for the final row scenario of \tabref{tab:results}. The sharp drop-off at the right-hand-side of the plot indicates that  generating more haplotypes would have a negligible effect on the overall log-likelihood.

\subsection{Deconvolution}

In the absence of persons of interest, a useful task is to try and identify the profiles of the contributors to the mixture. From the set of joint haplotypes generated by the likelihood maximization, we can find marginal haplotype distributions for each unknown. 
The probabilities for the ten highest probability 
haplotypes for each  contributor generated for the final row of \tabref{tab:results}
 are displayed in \tabref{tab:happrobs}.
For the major contributor
the haplotype with the highest probability, at 0.6792, coincides with known haplotype of person A.\footnote{A table with the corresponding haplotypes is in the supplementary material.}

For the middle contributor (final column U3), the  highest probability haplotype, with probability 0.6600, coincides with the 
haplotype of contributor D except on the locus  DYS385a/b; for which the profile 13/17 is predicted.
Looking again at the contributor profiles on DYS385a/b in \tabref{tab:profiles}, we see that the major contributor A has profile 13/18, of which neither allele is in the profiles of the other two contributors. The  peak heights for these alleles 
on this locus shown in \tabref{tab:peaks} shows a large imbalance, so it is perhaps not too surprising that allele 13 is also predicted to be in the profile of the middle contributor. The correct profile on this locus, 15/17, occurs in less than 0.3\% of the haplotypes generated compared to 75\% for the profile 13/17 (data not shown). The profile for C is predicted less well, but given the low amount of amplifiable DNA from this contributor expected to be in the replicate, this is not surprising. 

\begin{table}[htbp]
\caption{Ranked marginal haplotype profile probabilities for each untyped person.}
\begin{center}
\begin{tabular}{ccc}
U1 &U2 &U3 \\ \hline
0.6792 &0.1985 &0.6600 \\
0.0055 &0.0732 &0.0116 \\
0.0052 &0.0718 &0.0067 \\
0.0048 &0.0595 &0.0053 \\
0.0048 &0.0270 &0.0051 \\
0.0047 &0.0261 &0.0047 \\
0.0044 &0.0217 &0.0046 \\
0.0041 &0.0215 &0.0045 \\
0.0041 &0.0155 &0.0043 \\
0.0040 &0.0154 &0.0040 \\ \hline
\end{tabular}
\end{center}
\label{tab:happrobs}
\end{table}

\subsection{Assessing model fit}
\begin{figure}[hbtp]
\centering
\includegraphics[scale=0.6]{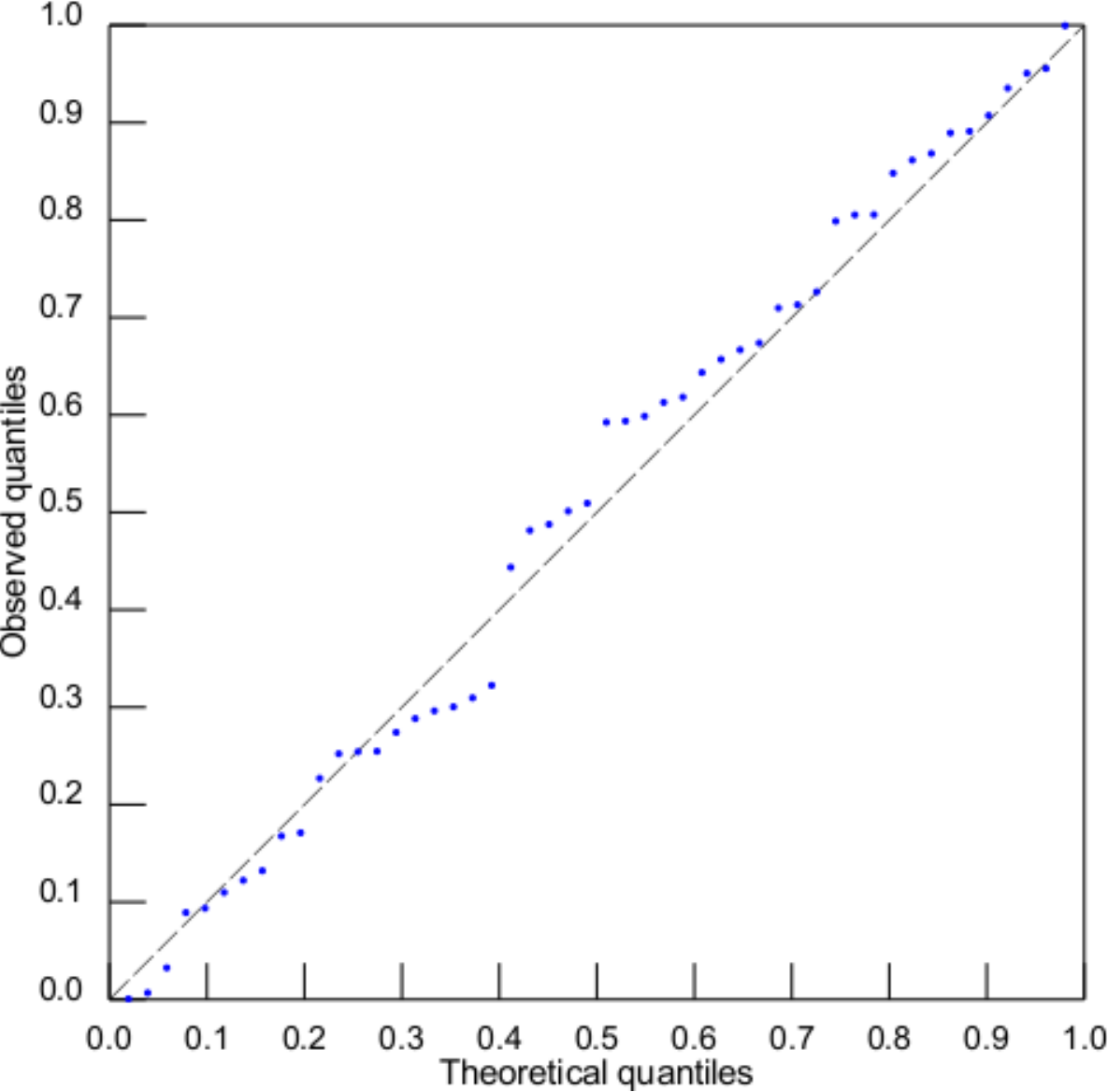}
\caption{Probability for the  ACD scenario in the Yfiler example for detected peaks at or above threshold. With an ordering of all alleles, the
observed quartile for an allele $a$ with peak height $h_a$ above the analytic threshold $T$ is 
$P(H_a \le h_a \cd H_a \ge T, \{h_b : b < a\})$. The points should follow the diagonal line.}
\label{fig:qqplotyf}
\end{figure}

Statistical tools for assessing the adequacy of the fit of statistical models of DNA mixtures were developed in \citep{graversen2014statistical}. One of these is a probability plot 
generated for peaks observed at or above the analytic threshold. 
As emphasized in \citep{graversen2019yara}, by basing the evaluation on the prequential framework of \citep{seillier1993testing}, the points in the graph  are independent, and if the underlying probability model is correct they are are uniformly distributed, so that plotting observed quantiles against quantiles from a uniform distribution should yield a straight line of slope 1.
Such a plot for the 
ACD scenario in shown  in \figref{fig:qqplotyf}; the fit looks good.

\begin{landscape}
\begin{figure}[hbtp]
\centering
\includegraphics[scale=0.970]{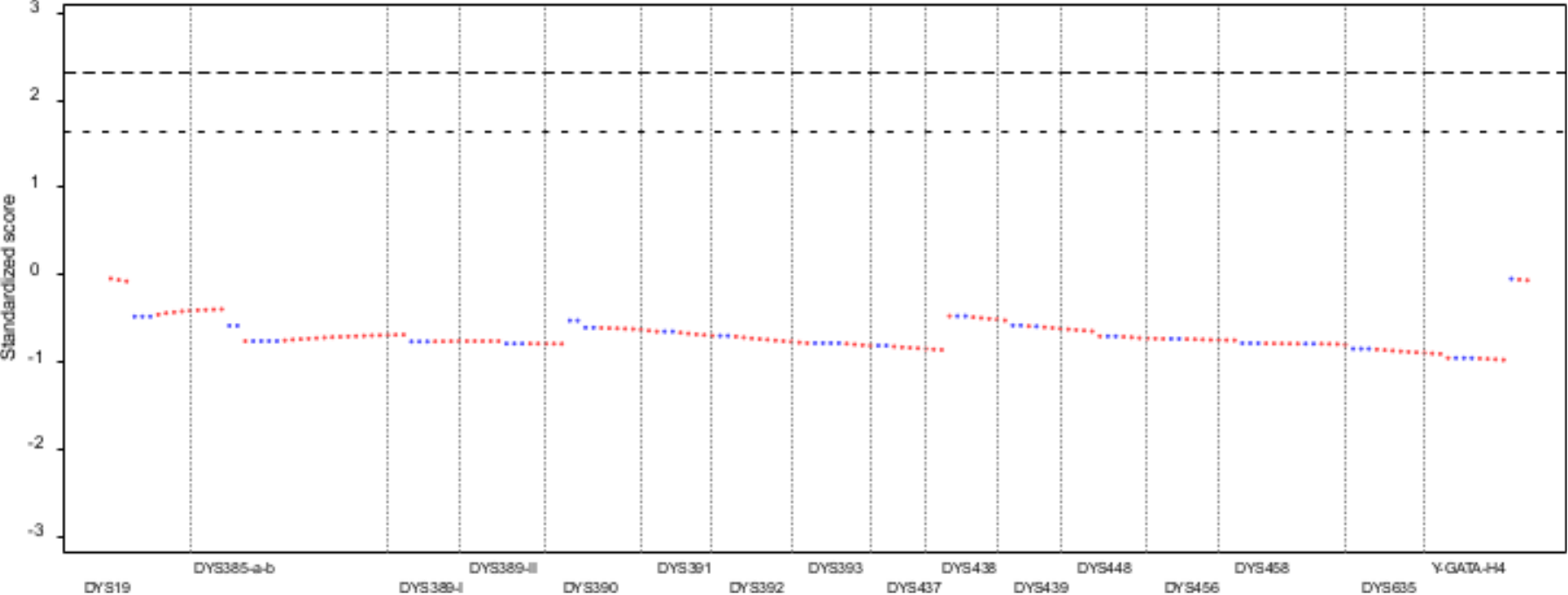}
\caption{Prequential monitor plot for the ACD scenario in the Yfiler example. 
Alleles with peaks at or above the analytic threshold are shown in blue, otherwise they are shown in red.  The two sets of broken lines are the upper 95\% and 99\% confidence limits. Ideally, if the model is adequately predicting the presence and absence of peaks, the end point of the plot should lie comfortably below these, which it does. The large upward jump in the YGATA-H4 locus arises from the peak height of 27 RFUs at the allele 11.1. This allele is not present in any of the three typed persons, so the jump in the plot could  be explained for this scenario either as high baseline noise at this peak or allelic dropin (or combination thereof); alternatively, and perhaps more likely in this case, it may be some other artefact occurring in the electropherogram data.
}
\label{fig:preqplotyf}
\end{figure}
\end{landscape}

\cite{graversen2014statistical} also introduced  prequential monitor plots for assessing the model predictions of whether or not a peak is seen
at or above the analytic threshold for each allele position.  \figref{fig:preqplotyf} shows such a plot for the ACD scenario, in which the scores have been normalized. Asymptotically the normalized score has a standard normal distribution if the model is correctly predicting the presence or absence of peaks. 
The final point of the plot  lies comfortably  below the 95\% confidence limit, confirming  a good fit to the data. 
For more detailed explanations, uses, and examples  of these
diagnostic plots see \citep{graversen2014statistical,graversen2015computational}.

\clearpage

\section{Discussion concerning the Y haplotype mixture model}

We now discuss a  number of issues arising from the Y haplotype mixture model presented in this paper.

\subsection{Sensitivity to parameter choices}
\label{sec:robust}

The results of the model presented in this paper will depend on the choices made for the haplotype generation parameters $k$ and $m$, and the pattern factors $a,b,c$ and $d$. This raises the questions of how the values of these parameters should be set, and how sensitive or robust are results to variations in these parameters. 

We begin with the parameters $k$ (the upper bound on the number of single locus joint profiles to be produced) and $m$ (the upper bound on the number of haplotypes to produce from the single locus profiles).  
The common answer to these questions is that it will depend on the problem at hand, and the simplest way to choose them is to examine the range of likelihoods or log-likelihoods they produce. For the parameter $k$, one can first maximise the likelihood using the product rule,  and then carry out a  deconvolution of the mixture to 
obtain a ranked ordering the joint profiles of the untyped persons on each locus. Thus, for the example in the paper, 
with $k=100$ the ratio (on each locus) of the highest probability joint profile to the lowest probability profile exceeds $10^{9}$, and for $k=1500$ (the value used in the evaluations in this paper) this rises to a value in excess of $10^{20}$ on every locus. It would thus seem from these figures that $k=1500$ is more than sufficient for the example in this paper.

Choosing the value for $m$, the number of haplotypes to generate, is based on similar considerations, by looking at the range of values  of the log-likelihoods of the haplotypes generated, and making a judgement call as to whether this is sufficient. 
Making such a judgement is aided by the production of a plot such as shown in \figref{fig:hapdist}.
That figure shows what appears a dramatic drop in values as the rank of the generated haplotypes approaches 500000, strongly suggesting that the remaining possible haplotypes will contributed a negligible amount to the log-likelihood. The values in \tabref{tab:results} for the scenario of three untyped persons shows a small change in the log-likelihoods and estimated cell counts when raising $m$ from 5000 to 500,000. Raising $m$ to 5 million yields a log-likelihood value of -115.427, and estimated cell counts of $(1559, 439, 1089)$.

The question of estimating the pattern factors  was addressed
\secref{sec:rationale}. It is a simple matter to vary these parameters to see the sensitivity to the resulting likelihoods and cell count estimates.
For example, on setting all the parameters to 0, the maximized log likelihood for the three untyped person scenario, with $m=5000$ haplotypes, is -116.366, a change of 0.004. The cell count estimate is $(1518, 449, 1123)$, again very close to the values in 
\tabref{tab:results}. On raising all parameters to 0.2, quite a bit higher than the estimates in \secref{sec:rationale}, the maximized log likelihood is -116.302, with cell count estimates of $(1520, 453, 1119)$. We thus see that, \textit{at least for the example in this paper}, results are quite robust to the parameters used.

\subsection{Use of likelihood ratios in Y-STR mixture problems}
\label{sec:ystrlike}

This paper has presented a model for evaluating likelihoods for hypotheses involving contributors to Y-STR mixtures. It uses a branching process model to find the distribution and expected number of matching haplotypes for use in evaluating the overall likelihood \eqref{eq:like2}.

This method is based on the simulation study of 
\cite{andersen2017convincing}, however that paper questions whether
likelihood ratios are appropriate to present as evidence in court, and instead advocates giving the expectation, or an upper quantile (for example 99\%), of the number of males with matching Y haplotype profiles to a person of interest.
As the authors show in their simulations, and explained by the branching process modelling of this paper, these numbers are insensitive to the size of the population if the population is relatively large. However, a jury might find it hard to know how to use such numbers if they are not given some guidance on the size of the population the males might be from. Another issue is that their method appears to be appropriate to good quality, preferable single source samples; issues of degradation, dropin and dropout raise difficulties in what may constitute a suspect's (say) profile matching a replicate profile. Additionally it cannot be used to evaluate the likelihood terms in \eqref{eq:like2}.

In contrast, the likelihood ratios obtained from likelihoods of scenarios, as shown for example in \tabref{tab:happrobs}, will depend
depend on the population size $\Omega$ which is used as a divisor. For example, on reducing the size of the population used in the example down to 150,000 from the 200 million used,  the log-likelihood ratio of  AUU vs UUU  drops from 5.941 to  2.969, a change of  2.972.  This is closely, but not exactly,  in line with the population change: $\log_{10}(2\times 10^8/1.5 \times 10^5) = 3.125$.  What this shows is that when reporting a likelihood ratio, the assumed population size  must also be reported. 

Note, in contrast,  that the product rule likelihoods and likelihood ratios  do not depend on the population size. This will also be the case for the Discrete-Laplace model \citep{andersen2013discrete}.

By dividing the population size by the likelihood ratio, one can find an estimate of the expected number of untyped persons in the population that could `stand-in' for the person of interest in the prosecution hypothesis in explaining the observed replicate data (note that `stand-in' does not necessarily mean being having an exactly  matching profile to the person of interest).
For  AUU vs UUU, this number is approximately 229 using the value in \tabref{tab:happrobs} and a population size of 200 million. Using a  population size of 150,000, the expected number is approximately 161.
For comparison, the expected number of matching males is approximately 26.4 for the
population size of 200 million, and 25.4 for the population size of 150,000
(using the sample size of 718 for the number of European-Americans used from the
\citep{purps2014global} dataset). The median value given in \citep{andersen2017convincing} is around 26, and the upper 95\% quantile is approximately 120. We thus see that the values found using the mixture model are somewhat  conservatively in favour of the defence compared to the figures advocated in \citep{andersen2017convincing}.

\subsection{Analysis of more examples}
\label{sec:biggerdata}

This paper has used just one example to illustrate the model it proposes. Although the diagnostic checks appear to show the model working well, clearly it is desirable to validate it on more examples, preferably using data in the public domain to enable  comparison to other methods that may be developed in the future by other researchers. Using the data from \cite{purps2014global} is  one option, but as described in \secref{sec:realdata}  some settings had to be assumed so that the data could be analysed, so that care would be needed in interpreting the results.

An additional option is to carry out a simulation analysis. 
There are two core issues to be answered here: (i) how is data to be simulated?, and (ii) what analyses would constitute an acceptable validation?

As regards simulating data, is it a simple matter to  simulate the
PCR process for specified contributor haplotypes and DNA amounts.
The numbers and contributors are readily varied, as is their DNA amounts, enabling comparison of fitted values to known inputs.  However, the specification of the contributor haplotypes is not so straightforward. Simply simulating locus profiles independently using marginal allele counts of loci from some population will not generate haplotypes with the required locus correlations. Instead, simulating a large population would be required similar to that described in \citep{andersen2017convincing}. However such a simulation would need to be extended to included deletions, duplications and partial repeats arising from mutations. Having generated a large population, one or more reference databases of various sized can be  sampled from it, and contributor profiles may also be sampled.

So the next question to answer, is what analyses would constitute a validation for the methods? This is not so easy to answer. Some things to check are that likelihoods decrease as DNA amounts decrease, that likelihood ratios for minor contributors are generally lower than for major contributors. Predictions of contributor amounts can be compared to  the amounts specified in the mixture simulations. 
These are essentially sanity checks that the model is performing as it is expected to. 

A more formal validation is $H_d$-true testing \citep{taylor2015testing} which has been applied to validating probabilistic genotyping systems for autosomal mixtures.
Basically, $H_d$-true testing involves taking a given defence hypothesis $H_d$ having one or more untyped persons, and evaluating its likelihood: denote this by $L_d$. One then repeatedly 
samples a profile from the population, $i$ say, and evaluates the
likelihood of the defence hypothesis $H_d$ with one of the  untyped persons replaced by the now known sampled profile; denote the likelihood by $L_i$.  One then computes the average of the likelihood ratios $ L_i/L_d$: in expectation this should be equal to 1 and the numerically obtained average can be compared to this.

\cite{taylor2015testing} and \cite{taylor2017importance} applied this to autosomal loci, with the latter reference using importance sampling to reduce the number of sampled profiles required (to enhance computational efficiency). However, to apply this to the Y-STR mixture cases requires some way to simulate a haplotype to be used for evaluating the $L_i$. As pointed out earlier, simply sampling a haplotype by sampling the loci  profiles individually will lead to the wrong statistics for the 
haplotype sampling probability. Additionally, the importance sampling weights used by \citep{taylor2017importance} were evaluated using the product rule, an option not available in the Y haplotype setting. 

One possible solution may be to sample from the haplotypes that have been simulated, or indeed just use all of them once. However the population will necessarily be finite, and it is not clear if this would be justified or would  biased the results (that is, it might happen that theoretically the expectation will not be 1). We thus leave it as an open question as to how to carry out 
 $H_d$-true testing in the Y haplotype setting.

\section{Summary}
\label{sec:summary}

A multivariate probability generating function analysis was presented for a sub-critical branching process model, which was inspired by the simulation study of \citep{andersen2017convincing} of Y haplotype diversity. The distributions obtained from the probability model appear to approximate very well  the Wright-Fisher simulations in  \citep{andersen2017convincing}.

A generic model for the evaluation of the peak height likelihoods for  Y haplotype samples has also been proposed. It involves generating a large set of high likelihood candidate haplotypes by initially treating the Y-STR loci as if they were independent. The product-rule haplotype probabilities are then replaced by  better haplotype probability estimates, and the  likelihood 
maximized over the candidate haplotypes weighted by these refined probabilities. A particular instance  for Y haplotype match probability has been proposed, 
based on the branching process model developed in this paper. 
 The haplotype probability model incorporates many of the 
complications known to arise
in Y haplotypes:  deletions, duplications and partial allelic repeats. 
The model is simple and intuitive.
An application to a  Yfiler profile from a DNA mixture of three males was presented which showed likelihoods much less extreme than those evaluated assuming the product rule. A deconvolution of the mixture, assumed to have three untyped males as contributors, into individual haplotypes for the untyped males was also presented, in which the major contributor was identified with the highest probability. Diagnostic probability and prequential plots were presented that indicate the model fitting the data well.

At present, there is no other model that can evaluate likelihoods taking into account the presence of deletions, duplications or partial repeats, and so a computational comparison to other methods was not possible.  A more systematic and extensive study is planned to see how the new model performs on real and simulated Y haplotype mixtures.

\subsection*{Acknowledgements}

I would like to thank Tim Clayton and Jim Thomson for enlightening me on the
special issues and concerns that distinguish the analysis of Y-STR haplotype profiles from the analysis of autosomal
profiles. 

I would also like to thank  anonymous referees for their comments on an earlier draft of this paper.


\begin{thebibliography}{26}
\providecommand{\natexlab}[1]{#1}
\providecommand{\url}[1]{\texttt{#1}}
\expandafter\ifx\csname urlstyle\endcsname\relax
  \providecommand{\doi}[1]{doi: #1}\else
  \providecommand{\doi}{doi: \begingroup \urlstyle{rm}\Url}\fi

\bibitem[Andersen and Balding(2017)]{andersen2017convincing}
M.~M. Andersen and D.~J. Balding.
\newblock How convincing is a matching {Y}-chromosome profile?
\newblock \emph{PLoS genetics}, 13\penalty0 (11):\penalty0 e1007028, 2017.

\bibitem[Andersen et~al.(2013{\natexlab{a}})Andersen, Caliebe, Jochens,
  Willuweit, and Krawczak]{andersen2013estimating}
M.~M. Andersen, A.~Caliebe, A.~Jochens, S.~Willuweit, and M.~Krawczak.
\newblock Estimating trace-suspect match probabilities for singleton {Y-STR}
  haplotypes using coalescent theory.
\newblock \emph{Forensic Science International: Genetics}, 7\penalty0
  (2):\penalty0 264--271, 2013{\natexlab{a}}.

\bibitem[Andersen et~al.(2013{\natexlab{b}})Andersen, Eriksen, and
  Morling]{andersen2013discrete}
M.~M. Andersen, P.~S. Eriksen, and N.~Morling.
\newblock The discrete {L}aplace exponential family and estimation of {Y-STR}
  haplotype frequencies.
\newblock \emph{Journal of Theoretical Biology}, 329:\penalty0 39--51,
  2013{\natexlab{b}}.

\bibitem[Andersen et~al.(2018)Andersen, Curran, de~Zoete, Taylor, and
  Buckleton]{andersen2018modelling}
M.~M. Andersen, J.~Curran, J.~de~Zoete, D.~Taylor, and J.~Buckleton.
\newblock Modelling the dependence structure of {Y-STR} haplotypes using
  graphical models.
\newblock \emph{Forensic Science International: Genetics}, 37:\penalty0 29--36,
  2018.

\bibitem[Bleka et~al.(2016)Bleka, Storvik, and Gill]{bleka2016euroformix}
{\O}.~Bleka, G.~Storvik, and P.~Gill.
\newblock {EuroForMix}: An open source software based on a continuous model to
  evaluate {STR DNA} profiles from a mixture of contributors with artefacts.
\newblock \emph{Forensic Science International: Genetics}, 21:\penalty0 35--44,
  2016.

\bibitem[Borel(1942)]{borel1942emploi}
{\'E}.~Borel.
\newblock Sur l’emploi du th{\'e}oreme de {B}ernoulli pour faciliter le
  calcul d’une infinit{\'e} de coefficients. application au probleme de
  l’attentea un guichet.
\newblock \emph{CR Acad. Sci. Paris}, 214:\penalty0 452--456, 1942.

\bibitem[Brenner(2010)]{brenner2010fundamental}
C.~H. Brenner.
\newblock Fundamental problem of forensic mathematics—the evidential value of
  a rare haplotype.
\newblock \emph{Forensic Science International: Genetics}, 4\penalty0
  (5):\penalty0 281--291, 2010.

\bibitem[Brenner(2014)]{brenner2014understanding}
C.~H. Brenner.
\newblock Understanding {Y} haplotype matching probability.
\newblock \emph{Forensic Science International: Genetics}, 8\penalty0
  (1):\penalty0 233--243, 2014.

\bibitem[Buckleton et~al.(2011)Buckleton, Krawczak, and
  Weir]{buckleton2011interpretation}
J.~S. Buckleton, M.~Krawczak, and B.~S. Weir.
\newblock The interpretation of lineage markers in forensic {DNA} testing.
\newblock \emph{Forensic Science International: Genetics}, 5\penalty0
  (2):\penalty0 78--83, 2011.

\bibitem[Butler(2011)]{butler2011advanced}
J.~M. Butler.
\newblock \emph{Advanced topics in forensic {DNA} typing: methodology}.
\newblock Academic Press, 2011.

\bibitem[Butler(2014)]{butler2014advanced}
J.~M. Butler.
\newblock \emph{Advanced topics in forensic {DNA} typing: interpretation}.
\newblock Academic Press, 2014.

\bibitem[Butler et~al.(2005)Butler, Decker, Kline, and
  Vallone]{butler2005chromosomal}
J.~M. Butler, A.~E. Decker, M.~C. Kline, and P.~M. Vallone.
\newblock Chromosomal duplications along the {Y}-chromosome and their potential
  impact on {Y-STR} interpretation.
\newblock \emph{Journal of Forensic Science}, 50\penalty0 (4):\penalty0
  JFS2004481--7, 2005.

\bibitem[Caliebe et~al.(2015)Caliebe, Jochens, Willuweit, Roewer, and
  Krawczak]{caliebe2015no}
A.~Caliebe, A.~Jochens, S.~Willuweit, L.~Roewer, and M.~Krawczak.
\newblock No shortcut solution to the problem of {Y-STR} match probability
  calculation.
\newblock \emph{Forensic Science International: Genetics}, 15:\penalty0 69--75,
  2015.

\bibitem[Cowell(2018)]{cowell2018unifying}
R.~G. Cowell.
\newblock A unifying framework for the modelling and analysis of {STR DNA}
  samples arising in forensic casework.
\newblock \emph{arXiv preprint arXiv:1802.09863}, 2018.

\bibitem[Cowell et~al.(1999)Cowell, Dawid, Lauritzen, and
  Spiegelhalter]{cowell:book}
R.~G. Cowell, P.~Dawid, S.~L. Lauritzen, and D.~J. Spiegelhalter.
\newblock \emph{Probabilistic networks and expert systems}.
\newblock Springer, 1999.

\bibitem[Graversen(2014)]{graversen2014statistical}
T.~Graversen.
\newblock \emph{Statistical and computational methodology for the analysis of
  forensic {DNA} mixtures with artefacts}.
\newblock PhD thesis, Oxford University, UK, 2014.

\bibitem[Graversen and Lauritzen(2015)]{graversen2015computational}
T.~Graversen and S.~Lauritzen.
\newblock Computational aspects of {DNA} mixture analysis.
\newblock \emph{Statistics and Computing}, 25\penalty0 (3):\penalty0 527--541,
  2015.

\bibitem[Graversen et~al.(2019)Graversen, Mortera, and Lago]{graversen2019yara}
T.~Graversen, J.~Mortera, and G.~Lago.
\newblock The {Y}ara {G}ambirasio case: Combining evidence in a complex {DNA}
  mixture case.
\newblock \emph{Forensic Science International: Genetics}, 40:\penalty0 52--63,
  2019.

\bibitem[Nilsson(1998)]{nilsson:97}
D.~Nilsson.
\newblock An efficient algorithm for finding the {$M$} most probable
  configurations in a probabilistic expert system.
\newblock \emph{Statistics and Computing}, 8:\penalty0 159--173, June 1998.

\bibitem[Purps et~al.(2014)Purps, Siegert, Willuweit, Nagy, Alves, Salazar,
  Angustia, Santos, Anslinger, Bayer, et~al.]{purps2014global}
J.~Purps, S.~Siegert, S.~Willuweit, M.~Nagy, C.~Alves, R.~Salazar, S.~M.
  Angustia, L.~H. Santos, K.~Anslinger, B.~Bayer, et~al.
\newblock A global analysis of {Y}-chromosomal haplotype diversity for 23 {STR}
  loci.
\newblock \emph{Forensic Science International: Genetics}, 12:\penalty0 12--23,
  2014.

\bibitem[Riley et~al.()Riley, Goor, and Hoffman]{osiris}
G.~Riley, R.~Goor, and D.~Hoffman.
\newblock Osiris: Free open-source {STR} analysis software.
\newblock www.ncbi.nlm.nih.gov/projects/SNP/osiris.
\newblock OSIRIS: Free Open-source STR Analysis Software, developed at the
  National Center for Biotechnology Information, NLM, NIH, Bethesda, MD.

\bibitem[Robin W.~Cotton and Terrill(2012)]{cotton:etal:2012}
C.~J.~W. Robin W.~Cotton, Catherine M.~Grgicak and M.~Terrill.
\newblock http://www.bu.edu/dnamixtures/, 2012.
\newblock Retrieved September 10, 2012.

\bibitem[Seillier-Moiseiwitsch and Dawid(1993)]{seillier1993testing}
F.~Seillier-Moiseiwitsch and A.~Dawid.
\newblock On testing the validity of sequential probability forecasts.
\newblock \emph{Journal of the American Statistical Association}, 88\penalty0
  (421):\penalty0 355--359, 1993.

\bibitem[Taylor et~al.(2015)Taylor, Buckleton, and Evett]{taylor2015testing}
D.~Taylor, J.~Buckleton, and I.~Evett.
\newblock Testing likelihood ratios produced from complex {DNA} profiles.
\newblock \emph{Forensic Science International: Genetics}, 16:\penalty0
  165--171, 2015.

\bibitem[Taylor et~al.(2017)Taylor, Curran, and
  Buckleton]{taylor2017importance}
D.~Taylor, J.~M. Curran, and J.~Buckleton.
\newblock Importance sampling allows {Hd} true tests of highly discriminating
  {DNA} profiles.
\newblock \emph{Forensic Science International: Genetics}, 27:\penalty0 74--81,
  2017.

\bibitem[Taylor et~al.(2018)Taylor, Curran, and
  Buckleton]{taylor2018likelihood}
D.~Taylor, J.~Curran, and J.~Buckleton.
\newblock Likelihood ratio development for mixed {Y-STR} profiles.
\newblock \emph{Forensic Science International: Genetics}, 35:\penalty0 82--96,
  2018.

\end{thebibliography}

\clearpage
\appendix
\section{Justification of convergence assumption}
\label{sec:converge}

A justification for the assumption of the convergence of the cluster size distribution may be found by considering a Wright-Fisher fixed size population of haplotypes. Let $N$ denote the population size. The generation of offspring may be simulated using a two step process:

\begin{enumerate}
\item Sample with replacement $N$ haplotypes.
\item Randomly mutate with probability $\mu$ each sampled haplotype; if mutation occurs the new haplotype is unique.
\end{enumerate}

Now for a fixed population size $N$ and generation $g$, the cluster sizes
$\tvec{c}_g = (c_{g,1}, c_{g,2}, \cdots )$ will have the constraint that
$$\sum_j jc_{g,j} = N.$$
Essentially, the clusters define a partition of the integer $N$. For any integer $N$ there is a finite set of partitions.
The evolution of the population can be viewed as Markov chain over this set of partitions.  It is irreducible because is it possible to move from one partition to any other partition in two generations. To see this, denote the
two generation clusters by $\tvec{c}_0$ and $\tvec{c}_2$, say. Let the intermediate generation haplotype consist of $N$ unique haplotypes, so that
$ \tvec{c}_1 = (N,0,0,\cdots )$. A transition from $\tvec{c}_0 $  to $\tvec{c}_1 $ is possible
by having all of the $N$ sampled haplotypes in Step~1 above mutate to new haplotypes in Step~2 with positive probability.

A transition from these unique haplotypes to haplotypes having the cluster sizes 
 $\tvec{c}_2 = (c_{2,1}, c_{2,2}, c_{2,3}, \cdots )$ defining the next generation is possible as follows.
Sample $c_{2,1}$ haplotypes from generation 1 each just once;
sample another  $c_{2,2}$ distinct haplotypes each just twice; sample another  $c_{2,3}$ distinct haplotypes each just three times; and so on. Then for Step~2 have that none of the haplotypes, save possibly for the singletons, mutate. The resulting set of haplotypes will then have the cluster partition  $\tvec{c}_2$ with positive probability.

Thus we have shown that the Markov chain is irreducible, and so has an equilibrium distribution to which it eventually converges.

In the paper we consider large populations that vary in size. It seems reasonable to assume, given the argument above for fixed population size, that the cluster sizes in a large population that has evolved over a long time to have approximately converged to a limiting distribution.

\end{document}